\documentclass[journal]{IEEEtran}
\usepackage{amsmath,amsfonts,amsthm}
\usepackage[ruled,commentsnumbered,linesnumbered]{algorithm2e}
\usepackage{array}
\usepackage{textcomp}
\usepackage[dvipsnames]{xcolor}
\usepackage{stfloats}
\ifCLASSOPTIONcompsoc
\usepackage[caption=false,font=normalsize,labelfont=sf,textfont=sf]{subfig}
\else
\usepackage[caption=false,font=footnotesize]{subfig}
\fi
\usepackage{url}
\usepackage{verbatim}
\usepackage{graphicx}
\usepackage{cite}
\usepackage[colorlinks,linkcolor=blue,citecolor=blue,urlcolor=blue]{hyperref} 
\usepackage{makecell}
\usepackage{multirow}
\usepackage{color}
\usepackage{booktabs}
\usepackage{amsmath}

\SetKwComment{Comment}{~$\triangleright$~}{}
\allowdisplaybreaks[4]
\usepackage{amssymb}
\usepackage{threeparttable}

\theoremstyle{remark}
\newtheorem{remark}{Remark}

\newtheorem{lemma}{Lemma}
\newtheorem{definition}{Definition}

\newtheorem{assumption}{Assumption}
\newtheorem{theorem}{Theorem}

\newtheorem{example}{Example}
\usepackage{orcidlink} 
\begin{document}

\title{Finite-Precision Conjugate Gradient Method for Massive MIMO Detection}
\author{Yiming Fang,~\IEEEmembership{Graduate Student Member,~IEEE}, 
    Li Chen,~\IEEEmembership{Senior Member,~IEEE},
    Changsheng You,~\IEEEmembership{Member,~IEEE},\\
    Dingzhu Wen,~\IEEEmembership{Member,~IEEE},
    and Pengcheng Zhu,~\IEEEmembership{Member,~IEEE}
   
    \thanks{Yiming Fang and Li Chen are with the CAS Key Laboratory of Wireless-Optical Communications, University of Science and Technology of China, Hefei 230027, China (e-mail: fym1219@mail.ustc.edu.cn; chenli87@ustc.edu.cn).}
    \thanks{Changsheng You is with the Department of Electronic and Electrical Engineering, Southern University of Science and Technology (SUSTech), Shenzhen, China (e-mail: youcs@sustech.edu.cn).}
    \thanks{Dingzhu Wen is with Network Intelligence Center, School of Information Science and Technology, ShanghaiTech University, Shanghai, China (e-mail: wendzh@shanghaitech.edu.cn).}
    \thanks{Pengcheng Zhu is with the National Mobile Communications Research Laboratory, Southeast University, Nanjing 210096, China (e-mail: p.zhu@seu.edu.cn).}
}



\maketitle
\begin{abstract}
The implementation of the conjugate gradient (CG) method for massive MIMO detection is computationally challenging, especially for a large number of users and correlated channels. In this paper, we propose a low computational complexity CG detection from a finite-precision perspective. First, we develop a finite-precision CG (FP-CG) detection to mitigate the computational bottleneck of each CG iteration and provide the attainable accuracy, convergence, and computational complexity analysis to reveal the impact of finite-precision arithmetic. A practical heuristic is presented to select suitable precisions. Then, to further reduce the number of iterations, we propose a joint finite-precision and block-Jacobi preconditioned CG (FP-BJ-CG) detection. The corresponding performance analysis is also provided. Finally, simulation results validate the theoretical insights and demonstrate the superiority of the proposed detection.
\end{abstract}

\begin{IEEEkeywords}
Block Jacob, conjugate gradient, finite-precision arithmetic, massive MIMO detection
\end{IEEEkeywords}

\section{Introduction}
\label{sec:intro}
\IEEEPARstart{M}{assive} multiple-input-multiple-output (MIMO) significantly improves the energy and spectral efficiencies for the next-generation wireless networks \cite{6457363}. Nevertheless, in practical implementation, the complexity of massive MIMO detection becomes prohibitive due to the substantial increase in system dimensions \cite{8804165}. Therefore, low-complexity massive MIMO detection has drawn significant attention.

To alleviate the hardware complexity of massive MIMO detection, existing literature mainly focuses on low-resolution analog-to-digital converters (ADCs) and novel analog computing architecture. Specifically, the authors in \cite{6987288} proposed an iterative multiuser detection based on a message-passing de-quantization algorithm for massive MIMO systems with low-resolution ADCs. Moreover, supervised and semi-supervised learning were applied in \cite{8959398} to enhance MIMO detection performance with low-resolution ADCs. Considering the analog computing architecture, the authors in \cite{10226413} implement zero forcing (ZF) based on resistive random-access memory with a gain of 91.2× in latency and 671× in energy efficiency. A fully parallel memristor-based circuit was proposed for linear minimum mean squared error (LMMSE) detection in \cite{10273392}. The authors in \cite{10738291} presented a memristor-based design for the optimal maximum likelihood (ML) MIMO detection.

In addition to the hardware complexity, the computational complexity of massive MIMO detection has also generated considerable interest. Even for the linear detection schemes like ZF and LMMSE, the increasing system dimension leads to complicated matrix inversion. To address this problem, various iterative MIMO detection methods have been proposed, which can be broadly classified into matrix-splitting (MS)/stationary methods and gradient-based/non-stationary methods. In MS methods, the Gram matrix is typically decomposed into its diagonal, upper, and lower triangular components. Specifically, the Jacobi iteration (JI) method \cite{7794623} and the Gauss-Seidel (GS) method \cite{8511052} used the inversion of the diagonal and lower triangular parts, respectively. As for the successive over-relaxation (SOR) method \cite{7399337}, the inversion of both the upper and lower triangular parts are incorporated to achieve faster convergence than the JI and GS methods. Moreover, the authors in \cite{9790338} proposed the randomized iterative detection algorithm that adopts random sampling into the MS methods.

Moreover, gradient-based iterative methods are promising to reduce the computational complexity of massive MIMO detection. The authors in \cite{10024792} employed gradient descent (GD) to implement iterative MIMO detection. Besides, the steepest descent (SD) method \cite{xue2018adaptive} and Barzilai-Bowein (BB) method \cite{8598346} were introduced to solve the problem of massive MIMO detection. The authors in \cite{9806305} leveraged limited-memory Broyden-Fletcher-Goldfarb-Shanno (L-BFGS) to massive MIMO detection, demonstrating convergence comparable to that of BFGS but with lower complexity. In addition, the conjugate gradient (CG) method exploits the Hermitian property of the Gram matrix to determine more efficient update directions \cite{7037382}, thereby accelerating convergence compared to GD, SD, and BB methods while achieving performance equivalent to L-BFGS.

The computational complexity of iterative detection, like CG detection, comes from two parts. First, it depends on the number of iterations, which tends to be high for correlated channels. To decrease the iteration number, classic preconditioning methods like Jacobi \cite{10187714}, incomplete Cholesky (IC) decomposition \cite{7868573}, and symmetric successive over relaxation (SSOR) \cite{9007506} have been used. Additionally, the authors in \cite{10653741} proposed user-wise singular value decomposition (UW-SVD) to accelerate the convergence of iterative detection. Second, the computational complexity is influenced by the computational cost of each CG iteration, which becomes prohibitive as the number of users increases. To address this issue, the recursive CG (RCG) detection and deep learning-based CG detection were introduced in \cite{8950336} and \cite{9250659}, respectively. Moreover, the authors in \cite{8114173} proposed decentralized CG (DCG) detection to reduce the computational complexity per iteration. 

The aforementioned works on low-complexity CG detection commonly implement the matrix computations using full-precision arithmetic ($\mathtt{fp64}$). Note that finite-precision or low-precision arithmetic offers further opportunities to reduce computational demands compared with full-precision arithmetic \cite{higham2002accuracy}. For instance, half-precision arithmetic provides advantages over double-precision arithmetic, including a fourfold increase in processing speed, a quarter of the storage requirements, and a $75\%$ reduction in memory transfer costs \cite{higham2022mixed}. Recently, finite-precision arithmetic has emerged as a powerful technique for reducing computational complexity in computer science and wireless communications. Specifically, large language models (LLMs), such as Deepseek \cite{liu2024deepseek,liu2024deepseek1} and LLaMA \cite{touvron2023llama,10884770}, frequently utilize finite-precision arithmetic to accelerate matrix computations. In the context of wireless communications, the authors in \cite{7168929} and \cite{9110827} proposed fixed-point arithmetic CG detection and finite-alphabet MIMO detection, respectively. Moreover,  the impact of finite-precision arithmetic on communication performance was analyzed, and a mixed-precision transceiver design was proposed in \cite{10845867}.

However, to the best of our knowledge, reducing the computational complexity of CG detection from a finite-precision perspective has never been studied. It incurs two challenges. First, the impact of errors caused by finite-precision arithmetic on CG detection is unknown. Second, how to reduce the corresponding errors to achieve near the performance of full-precision arithmetic is also an unresolved issue.

To fill in this gap, in this paper, we propose a low computational complexity CG detection from a finite-precision perspective. First, we propose a finite-precision CG (FP-CG) detection to alleviate the computational bottleneck of each CG iteration, and the attainable accuracy, convergence, and computational costs of FP-CG detection are presented. A heuristic method is provided to choose suitable precisions. Then, the joint finite-precision and block Jacob preconditioning CG (FP-BJ-CG) detection is proposed to reduce the iteration number based on the FP-CG detection, and the corresponding performance analysis is derived. Finally, simulation results show that FP-BJ-CG detection achieves only $1.2$ dB bit error rate (BER) performance loss at high signal-to-noise ratio (SNR) levels with decreasing $81.25\%$ of the computational cost compared to LMMSE detection. Our main contributions are summarized as follows. 
\begin{itemize}
    \item \textbf{Impact of finite-precision arithmetic for CG detection.} We propose FP-CG detection by employing finite-precision arithmetic to alleviate computational bottlenecks. Furthermore, the corresponding attainable accuracy is derived, which demonstrates the impact of precision, the dimension of the Gram matrix, and the condition number of the Gram matrix. More importantly, we reveal some intriguing observations. Specifically, using finite-precision for inner products in CG detection does not affect attainable accuracy; employing finite-precision for matrix-vector products in CG detection will significantly impact the attainable accuracy, particularly in ill-conditioned channel scenarios. Moreover, the convergence and computational complexity of FP-CG detection are also discussed.
    \item \textbf{Joint finite-precision and block Jacob preconditioning CG detection.} To further reduce the iteration number, we propose FP-BJ-CG detection. Specifically, considering correlated MIMO channels, we first prove theoretically that the Gram matrix tends to be near block diagonally dominant. Then, leveraging the near block diagonally dominant property, block Jacob (BJ) preconditioning is introduced to accelerate convergence with low computational complexity. Moreover, we provide a comprehensive performance analysis to show the superiority of FP-BJ-CG detection.
\end{itemize}

\textit{Organization:} Section \ref{sec: sys} describes a multi-user massive MIMO system model. In Section \ref{sec: fp-cg}, we introduce the preliminaries of finite-precision arithmetic, propose finite-precision CG detection, and analyze the impact of finite-precision arithmetic. Moreover, a practical heuristic is developed. In Section \ref{sec: joint}, the joint finite precision and preconditioning CG detection is proposed. Numerical results are presented in Section \ref{sec: sim}, and the conclusions are provided in Section \ref{sec: con}.

\textit{Notation:} Bold uppercase letters denote matrices, and bold lowercase letters denote vectors. For a matrix $\bf A$, ${\bf A}^\mathsf{T}$, ${\bf A}^\mathsf{H}$ and ${\bf A}^{-1}$ denote the transpose, the Hermitian transpose and inverse of ${\bf A}$, respectively. ${a}_{i,j}$ denotes $(i,j)$th entry of ${\bf A}$. $\mathrm{tr}(\bf A)$ denotes the trace of matrix ${\bf A}$. $\mathbb{E}\left(\cdot\right)$ and $\mathbb{V}\left(\cdot\right)$ denotes the expectation and variance, respectively. Given random variable $a$ and $b$, $\mathbb{C}(a,b)$ are the covariance between $a$ and $b$. $ \left| {\bf A}\right|$ represents the matrix of absolute values, $(\left| {a}_{i,j}\right|)$. $\left\| {\bf A} \right\|_2$ denotes its spectral norm. $\kappa({\bf A})=\left\| {\bf A} \right\|_2\left\| {\bf A}^{-1} \right\|_2$ represents the condition number of ${\bf A}$. For a vector $\bf a$, $\left\| \bf a \right\|_2$ denotes its Euclidean norm. The notations $\mathbb{R}$ and $\mathbb{C}$ represent the sets of real numbers and complex numbers, respectively. $\Re\{x\}$ and $\Im\{x\}$ denote the real part and imaginary part of $x$. $\lceil x \rceil$ and $\lfloor x \rfloor$ represent the smallest integer more than $x$ and the largest integer no more than $x$, respectively.

\section{System Model}
\label{sec: sys}
Consider an uplink multi-user massive MIMO system with a base station (BS) equipped with $M$ antennas, serving $K$ multi-antenna users indexed by $k\in \{1,2,\cdots, K\}$. The $k$-th user has $N_k$ antennas with $k\in \{1,2,\cdots, K\}$. Then, the received signals $\mathbf{y}\in \mathbb{C}^{M\times 1}$ at the BS can be expressed as
\begin{align}
    \mathbf{y=Hx+n},
\end{align}
where $\mathbf{H}=\left[ \mathbf{H}_1,\cdots, \mathbf{H}_K \right] \in \mathbb{C} ^{M\times N}$ is the channel matrix, $N=\sum_{k=1}^K N_k\ll M$, $\mathbf{H}_k \in \mathbb{C}^{M\times N_k}$ is the channel matrix between the BS and the $k$-th user, $\mathbf{x}\in \mathcal{CN}\left(0,{\bf I}_{N}\right)$ is the transmitted signals from all the users, $\mathbf{n}\in \mathcal{CN}\left(0,\sigma_n^2 {\bf I}_{M}\right)$ is the the additive white Gaussian noise (AWGN), and $\sigma_n^2$ is the noise variance.

Further, the channel matrix $\mathbf{H}$ is modeled by the correlated model as follows: \cite{951380}
\begin{align}\label{eq: channel_model}
    \mathbf{H}=\sqrt{\mathbf{R}_r}\mathbf{\Omega}\sqrt{\mathbf{R}_t},
\end{align}
where $\mathbf{\Omega}\in \mathbb{C}^{M\times N}$ is i.i.d. complex Gaussian matrix with its $(i, j)$-th element stratifying $\omega_{ij} \sim \mathcal{CN}(0, 1/M)$, and $1/M$ denotes the normalized variance of each channel element \cite{10653741}. Both $\mathbf{R}_r \in \mathbb{C}^{M\times M}$ and $\mathbf{R}_t \in \mathbb{C}^{N\times N}$ are the exponential correlation matrices of the BS and users side, whose elements $r_{ij}$ is given by
\begin{align}
    r_{ij}=\begin{cases}
	\zeta ^{j-i}&		i\le j\\
	r_{ji}^{*}&		i>j\\
    \end{cases},
\end{align}
where $\zeta$ is the correlation coefficient with $0\leq \zeta < 1$.

Then, to recover the transmitted signals, classic linear detection like ZF and LMMSE is employed, which can be written as \cite{yang2015fifty}
\begin{align}\label{eq: inversion}
    \hat{\mathbf{x}} = {\bf A}^{-1}{\bf b},
\end{align}
where $\hat{\mathbf{x}}$ the estimated transmitted signals, ${\bf b}={\bf H}^\mathsf{H}{\bf y}$, and ${\bf A}\in\mathbb{C}^{N\times N}$ is Gram matrix. For ZF and LMMSE detection, $\bf A$ is given by
\begin{align}
    {\bf A}_{\rm ZF} &= {\bf H}^\mathsf{H}{\bf H}, \\
    {\bf A}_{\rm LMMSE} &= {\bf H}^\mathsf{H}{\bf H} + \sigma_n^2 {\bf I}_{N}.
\end{align}
Nevertheless, ZF and LMMSE detection involves matrix inversion of $\bf A$ leading to the computational complexity of $\mathcal{O}\left(N^3\right)$, which is unbearable for massive MIMO systems.

To avoid matrix inversion, \eqref{eq: inversion} can be seen as the solution of the linear system ${\bf A}\hat{\mathbf{x}}={\bf b}$. The CG method is a popular iterative algorithm to solve the linear system due to its fast convergence and low computational complexity, which has gained high attention in MIMO detection \cite{7037382,7868573,9250659,10551475}. Specifically, the iterations of the CG method can be described as
\begin{align}\label{eq: update_of_x}
    {\bf x}_i = {\bf x}_{i-1} + \alpha_{i-1}{\bf p}_{i-1},~i=1,2,\cdots
\end{align}
where $i$ is the iteration index, and $\alpha_{i-1}$ is the step size satisfying
\begin{align}\label{eq: alpha}
    \alpha_{i-1} = \frac{{{\bf r}_{i-1}^{\mathsf{H}}{\bf r}_{i-1}}}{{\bf r}_{i-1}^{\mathsf{H}}{\bf A}{\bf p}_{i-1}},
\end{align}
and ${\bf r}_i$ is the residual of the linear equation as follows
\begin{align}
    {\bf r}_i &= {\bf r}_{i-1} - \alpha_{i-1}{\bf A}{\bf p}_{i-1} \label{eq: update_of_r},
\end{align}
and ${\bf p}_{i-1}$ the update direction as follows
\begin{align}
    {\bf p}_i &= {\bf r}_i + \beta_{i}{\bf p}_{i-1} \label{eq: update_of_p},   
\end{align}
where $ \beta_{i}$ is a scalar parameter satisfying
\begin{align}\label{eq: beta}
    \beta_{i} = \frac{{\bf r}_i^{\mathsf{H}}{\bf r}_i}{{\bf r}_{i-1}^{\mathsf{H}}{\bf r}_{i-1}}.
\end{align}

\begin{algorithm}[t]
\label{alg: CG}
    \DontPrintSemicolon
    \SetAlgoNlRelativeSize{-1}
    \caption{CG Detection}
    \KwIn{${\bf A}, {\bf b}$, and the number of iteration $\mathcal{I}$}
    \KwOut{Estimated transmit signal vector $\hat{\bf x}$}

    ${\bf r}_0 = {\bf b}$, ${\bf p}_0 = {\bf r}_0$, $\hat{\bf x}_0 = {\bf 0}$
    
   \For{$i = 1:\mathcal{I}$}{

    Update $\alpha_{i-1}$ using \eqref{eq: alpha}

    Update ${\bf x}_i$ using \eqref{eq: update_of_x}

    Update ${\bf r}_i$ using \eqref{eq: update_of_r}

    Update $\beta_{i}$ using \eqref{eq: beta}

    Update ${\bf p}_i$ using \eqref{eq: update_of_p}
   
   }

   $\hat{\bf x} = {\bf x}_\mathcal{I}$
    
\end{algorithm}

To summarize, CG detection is detailed in \textit{Algorithm} \ref{alg: CG}. The total computational complexity of CG detection is $\mathcal{O}\left(\mathcal{I}N^2\right)$, where $\mathcal{I}$ is the iteration number. The implementation of CG detection remains computationally challenging due to two key factors. First, the primary computational burden in each iteration stems from the matrix-vector products ${\bf Ap}_{i-1}$, resulting in $\mathcal{O}\left(N^2\right)$ operations per iteration, which constitutes the main computational bottleneck, especially for large numbers of users. Second, CG detection requires large iteration numbers when dealing with ill-conditioned channels (e.g., highly correlated channels \cite{10653741,9007506}) due to the large condition number. Furthermore, the following example illustrates this challenge.

\begin{example}\label{ex: 1}
Consider a multi-user MIMO (MU-MIMO) system using full-precision arithmetic ($\mathtt{fp64}$), where the BS is equipped with a large number of antennas, e.g., $M = 512$, serving substantial single-antenna users, e.g., $N = K = 100$. According to the Monte-Carlo simulation, CG detection requires approximately $28$ iterations to achieve the BER performance of LMMSE detection at a high SNR level (SNR $=25$ dB) and under highly correlated channels ($\zeta = 0.8$). This corresponds to an estimated computational cost of approximately 280,000 floating-point operations (flops) per subchannel. Further, considering the orthogonal frequency-division multiplexing (OFDM) system supporting up to $3300$ subcarriers with a slot duration of $30\mathrm{\mu s}$ \cite{8412469}, the required computational cost for CG detection is on the order of $30$ tera floating-point operations per second (TFLOPS). Notably, the NVIDIA A100 graphics processing unit (GPU) can deliver over $9.7$ TFLOPS \cite{nvidia}. Consequently, multiple GPUs would be required to meet the computational demand, leading to an impractical increase in computational cost and power consumption.
\end{example}
Compared with full-precision arithmetic, finite-precision or low-precision arithmetic significantly enhances computational speed while also reducing storage and memory transfer costs. This naturally raises the question: \textit{Can finite-precision arithmetic be leveraged to reduce the computational complexity of CG detection?} To this end, we first employ finite-precision arithmetic to alleviate computational bottlenecks from matrix-vector products and then decrease the number of iterations by joint finite-precision and preconditioning CG detection.

\section{Finite-Precision CG Detection}
\label{sec: fp-cg}
In this section, we propose an FP-CG detection to mitigate computational bottlenecks in \textit{Algorithm} \ref{alg: CG}. First, we introduce the detailed FP-CG detection. Then, the performance of FP-CG detection is analyzed to evaluate the impact of finite-precision arithmetic, and we also give a heuristic method to show how to choose different precisions in different channel conditions without significantly affecting performance.

\subsection{Proposed FP-CG Detection}
Before introducing the proposed FP-CG detection, we first present essential definitions of basic finite-precision arithmetic below.
\begin{definition}[\textit{Finite-Precision Operator}]
\label{def: fp_op}
    $\boldsymbol{fl}\left(\cdot\right)$ is the operator that rounds a real number into the finite-precision floating-point number system $\mathbb{F}$.
\end{definition}

\begin{table}[t]
\centering 
\setlength{\tabcolsep}{3.5pt}
\caption{Parameters for Four Finite-Precision Arithmetic}
\label{tab: parafp}
\begin{threeparttable}
\begin{tabular}{cclll}
\toprule[1pt]
\midrule
 & $\left(\mathrm{sig.}, \mathrm{exp.}\right)$\tnote{(1)} & \multicolumn{1}{c}{$u$\tnote{(2)}} & \multicolumn{1}{c}{$x_{\min}$\tnote{(3)}} & \multicolumn{1}{c}{$x_{\max}$\tnote{(4)}} \\ \midrule
$\mathtt{bfloat16}$    &  $(8,8)$    & $3.91\times 10^{-3}$ & $1.18\times 10^{-38}$& $3.39\times 10^{38}$ \\ 
$\mathtt{fp16}$      &  $(11,5)$    & $4.88\times 10^{-4}$ & $6.10\times 10^{-5}$& $6.55\times 10^{4}$ \\ 
$\mathtt{fp32}$     &  $(24,8)$    & $5.96\times 10^{-8}$ & $1.18\times 10^{-38}$& $3.40\times 10^{38}$ \\ 
$\mathtt{fp64}$     &  $(53,11)$    & $4.88\times 10^{-16}$ & $2.22\times 10^{-308}$& $1.80\times 10^{308}$ \\ \midrule
\bottomrule[1pt]
\end{tabular}
    \begin{tablenotes}    
        \footnotesize               
        \item[(1)] $\left(\mathrm{sig.}, \mathrm{exp.}\right)$ represents number of bits in significand and exponent.          
        \item[(2)] $u$ is unit roundoff.
        \item[(3)] $x_{\min}$ is smallest normalized positive number.
        \item[(4)] $x_{\max}$ is largest finite number.
    \end{tablenotes}            
\end{threeparttable}
\end{table}

\begin{definition}[\textit{Standard Arithmetic Model}]
\label{def: samodel}
    Denote by $u$ the \textit{unit roundoff}. The floating-point system $\mathbb{F}$ adheres to a standard arithmetic model if, for any $x, y \in \mathbb{R}$ is in the range of $\mathbb{F}$. One has{\cite[Sec. 2.2]{higham2002accuracy}}
    \begin{equation}
    \label{eq: model}
        \boldsymbol{fl}\left(x\,\mathrm{op}\,y\right) = \left(x\,\mathrm{op}\,y\right)(1+\delta),
    \end{equation}
    where $\delta \in \mathbb{R}$ is such that $\left|\delta\right|\leq u$, for $\mathrm{op}=+,-,\times,/$.
\end{definition}
The standard arithmetic model in \textit{Definition} \ref{def: samodel} is satisfied by the IEEE arithmetic standard \cite{higham2022mixed}. Various \textit{unit roundoffs} $u$ in Table \ref{tab: parafp} can be chosen to represent different precisions.

Then, based on the analysis in Section \ref{sec: sys}, the computational bottleneck of each CG iteration is the matrix-vector products ${\bf Ap}_{i-1}$. To address this, we replace the matrix-vector products with finite-precision arithmetic, as summarized in \textit{Algorithm} \ref{alg: FP-CG}. For a more general analysis, inner products in each iteration are also considered to be computed using finite-precision arithmetic. 

\begin{algorithm}[t]
    \DontPrintSemicolon
    \SetAlgoNlRelativeSize{-1}
    \label{alg: FP-CG}
    \caption{FP-CG Detection with three precisions $u_{\rm MV}$, $u_{\rm IP}$ and $u_{\rm W}$}
    \KwIn{${\bf A}$, ${\bf b}$, and the number of iteration $\mathcal{I}$}
    \KwOut{Estimated transmit signal vector $\hat{\bf x}$}

    ${\bf r}_0 = {\bf b}$, ${\bf p}_0 = {\bf r}_0$, ${\bf x}_0 = {\bf 0}$

    \For{$i = 1:\mathcal{I}$}{

     \textcolor{red!60}{$\boldsymbol{ \varpi}_{i-1} = {\bf A}{\bf p}_{i-1}$ \Comment*[r]{$u_{\rm MV}$}} 

    ${\varphi}_{i-1} = {\bf p}_{i-1}^{\mathsf{H}}\boldsymbol{ \varpi}_{i-1}$ \Comment*[r]{$u_{\rm IP}$} 

    ${\varrho}_{i-1} = {{\bf r}_{i-1}^{\mathsf{H}}{\bf r}_{i-1}}$ \Comment*[r]{$u_{\rm IP}$}

     $\alpha_{i-1} ={\varrho}_{i-1} /{{\varphi}_{i-1}}$  \Comment*[r]{$u_{\rm W}$}

     ${\bf x}_i = {\bf x}_{i-1} + \alpha_{i-1}{\bf p}_{i-1} $ \Comment*[r]{$u_{\rm W}$}

    ${\bf r}_i = {\bf r}_{i-1} - \alpha_{i-1}\boldsymbol{ \varpi}_{i-1} $ \Comment*[r]{$u_{\rm W}$}

   ${\varrho}_{i} = {\bf r}_i^{\mathsf{H}}{\bf r}_i$ \Comment*[r]{$u_{\rm IP}$} 

    $\beta_{i} = {\varrho}_{i}/{{\varrho}_{i-1}}$ \Comment*[r]{$u_{\rm W}$}

    ${\bf p}_i = {\bf r}_i + \beta_{i}{\bf p}_{i-1}$ \Comment*[r]{$u_{\rm W}$}
    
    }

     $\hat{\bf x} = {\bf x}_{\mathcal{I}}$
    
\end{algorithm}

More specifically, \textit{Algorithm} \ref{alg: FP-CG} assumes that all the computations and data storage use full precision ($\mathtt{fp64}$), denoted as $u_{\rm W}$, except for matrix-vector products (\textit{Step 1}) and inner products (\textit{Steps 4, 5, and 9}), which are computed in precision $u_{\rm MV}$ and $u_{\rm IP}$, respectively. Furthermore, we assume that $u_{\rm W} \ll u_{\rm MV}=u_{\rm IP}$ to ensure negligible error when storing results from matrix-vector and inner product operations. This method effectively mitigates computational bottlenecks. Nevertheless, the impact of finite-precision arithmetic on CG detection performance remains unexplored. In the following subsection, we provide the performance analysis of FP-CG detection to evaluate the impact of finite-precision arithmetic.

\subsection{Performance Analysis of FP-CG Detection}
In this subsection, we analyze the performance of FP-CG detection and discuss the impact of finite-precision arithmetic. We begin by presenting key lemmas and theorems related to finite-precision arithmetic, which form the foundation for our performance analysis. Specifically, based on the model in \textit{Definition} \ref{def: samodel}, the standard results for classic matrix computations are given in the following lemma.

\begin{lemma}[\textit{Real-valued Matrix Computations \cite{higham2002accuracy,greenbaum1997estimating}}]
\label{lem: rv}
    Given $a \in \mathbb{R}$, $\mathbf{v,w}\in \mathbb{R}^{n\times 1}$, ${\bf B}\in \mathbb{R}^{m\times n}$ with $\mathrm{rank}\left( \mathbf{B} \right) = n$, and computed precision $u_s$, the following error bounds hold:
    \begin{align*}
        \left\| \boldsymbol{fl}\left( a\mathbf{v} \right) -a\mathbf{v} \right\| _2 &\le u_s\left\| a\mathbf{v} \right\| _2,\\
        \left\| \boldsymbol{fl}\left( \mathbf{v}+\mathbf{w} \right) -\left( \mathbf{v}+\mathbf{w} \right) \right\| _2&\le u_s\left( \left\| \mathbf{v} \right\| _2+\left\| \mathbf{w} \right\| _2 \right), \\
        \left\| \boldsymbol{fl}\left( \mathbf{v}^{\mathsf{T}}\mathbf{w} \right) -\mathbf{v}^{\mathsf{T}}\mathbf{w} \right\| _2&\le \gamma _{n}^{(s)}\left\| \mathbf{v} \right\| _2\left\| \mathbf{w} \right\| _2,\\
        \left\| \boldsymbol{fl}\left( \mathbf{Bv} \right) -\mathbf{Bv} \right\| _2&\le \sqrt{n}\gamma _{n}^{(s)}\left\| \mathbf{B} \right\| _2\left\| \mathbf{v} \right\| _2,
    \end{align*}
    where 
    \begin{align}\label{eq: gamma}
        \gamma _{n}^{\left( s \right)}=\frac{nu_s}{1-nu_s}=nu_s+\mathcal{O} \left( u_{s}^{2} \right)~\left( nu_s<1 \right). 
    \end{align}
\end{lemma}

Further, since the channel matrix and other communication parameters are modeled as complex matrices and vectors in Section \ref{sec: sys}, we extend the error bounds in \textit{Lemma} \ref{lem: rv} to encompass complex-valued arithmetic in the theorem below.

\begin{theorem}[\textit{Complex-valued Matrix Computations}]
\label{the: cv}
    Given $a \in \mathbb{C}$, $\mathbf{v,w}\in \mathbb{C}^{n\times 1}$, ${\bf B}\in \mathbb{C}^{m\times n}$ with $\mathrm{rank}\left( \mathbf{B} \right) = n$, and computed precision $u_s$, the following error bounds hold:
    \begin{align*}
        \left\| \boldsymbol{fl}\left( a\mathbf{v} \right) -a\mathbf{v} \right\| _2&\le \sqrt{2}\gamma _2^{(s)}\left\| a\mathbf{v} \right\| _2,\\
        \left\| \boldsymbol{fl}\left( \mathbf{v}+\mathbf{w} \right) -\left( \mathbf{v}+\mathbf{w} \right) \right\| _2&\le u_s\left( \left\| \mathbf{v} \right\| _2+\left\| \mathbf{w} \right\| _2 \right), \\
        \left\| \boldsymbol{fl}\left( \mathbf{v}^{\mathsf{H}}\mathbf{w} \right) -\mathbf{v}^{\mathsf{H}}\mathbf{w} \right\| _2&\le \sqrt{2}\gamma _{2n}^{(s)}\left\| \mathbf{v} \right\| _2\left\| \mathbf{w} \right\| _2,\\
        \left\| \boldsymbol{fl}\left( \mathbf{Bv} \right) -\mathbf{Bv} \right\| _2&\le \sqrt{2n}\gamma _{2n}^{(s)}\left\| \mathbf{B} \right\| _2\left\| \mathbf{v} \right\| _2.
    \end{align*}
\end{theorem}
\begin{IEEEproof}
    The proof is similar to that of {\cite[Theorem 1 \& 2]{10845867}}, which is omitted for conciseness.
\end{IEEEproof}

Then, based on \textit{Theorem} \ref{the: cv}, we can analyze the maximum attainable accuracy of FP-CG detection in \textit{Algorithm} \ref{alg: FP-CG}. Due to finite-precision arithmetic, the \textit{true} residuals ${\bf b-Ax}_i$ and the \textit{computed} residuals vectors ${\bf r}_i$ in \textit{Algorithm} \ref{alg: FP-CG} have a gap. Following the work of \cite{greenbaum1997estimating}, we can use the difference between ${\bf b-Ax}_i$ and ${\bf r}_i$ to analyze the maximum attainable accuracy of \textit{Algorithm} \ref{alg: FP-CG}, which is derived in the following theorem.

\begin{theorem}[\textit{Attainable Accuracy of \textit{Algorithm} \ref{alg: FP-CG}}]
\label{the: Attainable}
    The different with \textit{true} residuals ${\bf b-Ax}_i$ and the \textit{computed} residuals vectors ${\bf r}_i$ in \textit{Algorithm} \ref{alg: FP-CG}, i.e., attainable accuracy, satisfies
    \begin{align}
        \frac{\left\| \mathbf{b}-\mathbf{Ax}_i-\mathbf{r}_i \right\| _2}{\left\| \mathbf{A} \right\| _2\left\| \mathbf{x} \right\| _2}\le \left( 1+i \right) u_{\mathrm{W}}+ \varUpsilon_i\Theta _i, \label{eq: mp}
    \end{align}
    where 
    \begin{align}
        \varUpsilon_i = &\left( \left( 8\sqrt{2}+6 \right) i+1 \right) u_{\mathrm{W}} \notag \\
        &+\left( 2i+1 \right) 2\sqrt{2}N^{\frac{3}{2}}u_{\mathrm{MV}}, \\
        \Theta _i&=\max_{k\le i} \frac{\left\| \mathbf{x}_k \right\| _2}{\left\| \mathbf{x} \right\| _2}.
    \end{align}
    Further, provided that \textit{Algorithm} \ref{alg: FP-CG} converges, the attainable accuracy satisfies
    \begin{align}
        \frac{\left\| \mathbf{b}-\mathbf{Ax}_i-\mathbf{r}_i \right\| _2}{\left\| \mathbf{A} \right\| _2\left\| \mathbf{x} \right\| _2}&\le \left( 1+i \right) u_{\mathrm{W}}+\varUpsilon_i \left( 1+\sqrt{\kappa \left( \mathbf{A} \right)} \right) \\
        &\sim \mathcal{O} \left( N^{\frac{3}{2}}u_{\mathrm{MV}}\sqrt{\kappa \left( \mathbf{A} \right)} \right). \label{eq: or_ph}
    \end{align}
\end{theorem}
\begin{IEEEproof}
    The proof is available in Appendix \ref{app: attainable_accuracy}.
\end{IEEEproof}

\textit{Theorem} \ref{the: Attainable} reveals that the attainable accuracy of \textit{Algorithm} \ref{alg: FP-CG} depends on work precision $u_{\rm W}$, matrix-vector products precision $u_{\rm MV}$, the dimension of ${\bf A}$, and condition number $\kappa \left( \mathbf{A} \right)$. Under real-valued computations with $u_{\rm W}=u_{\rm MV}=u_{\rm IP}$, \eqref{eq: mp} reduces to the special case of real-valued CG using uniform precision shown in {\cite[Eq. (24)]{greenbaum1997estimating}}. The insights from \textit{Theorem} \ref{the: Attainable} are further highlighted in the following remark.

\begin{remark}[\textit{Impact of Finite-Precision Arithmetic on Attainable Accuracy}]\label{rem: impact_fp}
    \textit{Theorem} \ref{the: Attainable} highlights two key insights. First, since \eqref{eq: mp} and \eqref{eq: or_ph} do not involve the inner products precision $u_{\rm IP}$, $u_{\rm IP}$ has no influence on the residual gap and thus does not affect attainable accuracy of \textit{Algorithm} \ref{alg: FP-CG}. Second, given $u_{\rm W} \ll u_{\rm MV}$, the rounding errors introduced by matrix-vector products dominate the residual gap, which will significantly affect the attainable accuracy of \textit{Algorithm} \ref{alg: FP-CG} when low-precision arithmetic is used, particularly in ill-conditioned channel scenarios.
\end{remark}

\begin{remark}[\textit{Impact of Finite-Precision Arithmetic on Convergence}]\label{rem: con}
    In \textit{Theorem} \ref{the: Attainable}, \eqref{eq: mp} is derived under the assumption of convergence, and hence we need to discuss the convergence properties of \textit{Algorithm} \ref{alg: FP-CG}. On the one hand, using finite-precision arithmetic, particularly low-precision arithmetic, may delay convergence due to the loss of local orthogonality \cite{greenbaum2021convergence}. On the other hand, applying the conclusion from \cite[Theorem 5.24]{meurant2006lanczos} to \textit{Algorithm} \ref{alg: FP-CG} reveals that if $u_{\mathrm{MV}}\left\| \mathbf{A}^{-1} \right\| _2\ll 1$, convergence is guaranteed. 
\end{remark}

As discussed in \textit{Remarks} \ref{rem: impact_fp} and \ref{rem: con}, employing finite-precision arithmetic, especially low-precision arithmetic, can negatively impact the attainable accuracy and convergence of CG detection. This raises a critical question: \textit{How should one select precisions to avoid significant degradation in CG detection performance?} To this end, we expect that
\begin{align}
    \frac{\left\| \mathbf{b}-\mathbf{Ax}_i-\mathbf{r}_i \right\| _2}{\left\| \mathbf{A} \right\| _2\left\| \mathbf{x} \right\| _2} \ll 1,
\end{align}
and using \eqref{eq: or_ph} yields the condition
\begin{align}
    \mathcal{O} \left( N^{\frac{3}{2}}u_{\mathrm{MV}}\sqrt{\kappa \left( \mathbf{A} \right)} \right) \ll 1.
\end{align}
Since the derivation of \eqref{eq: or_ph} is based on \textit{Theorem} \ref{the: cv}, which provides worst-case bounds, these bounds tend to be conservative and overestimated \cite{higham2019new}. Therefore, we can roughly omit $\mathcal{O}(\cdot)$, replace $\ll$ with $<$, and formulate the \textit{heuristic} method:
\begin{align}
    N^{1.5}u_{\mathrm{MV}}\sqrt{\kappa \left( \mathbf{A} \right)}<1\Longrightarrow u_{\mathrm{MV}}<\frac{1}{N^{1.5}\sqrt{\kappa \left( \mathbf{A} \right)}} \label{eq: heuristic}.
\end{align}

\textit{In summary, given $N$ and $\kappa(\bf A)$\footnote{{In practice, instead of the instantaneous condition number, one can employ the statistical expectation $\kappa(\bf A)$, i.e., $\mathbb{E}\{\kappa(\bf A)\}$, which has been investigated in \cite{5474635,7790799,0609045}.}}, we can first compute the upper bound of $u_{\mathrm{MV}}$ based on \eqref{eq: heuristic} and then consult Table \ref{tab: parafp} to select suitable precision corresponding to values of $u_{\mathrm{MV}}$.} In Section \ref{sec: sim}, simulation results will demonstrate that, while \eqref{eq: heuristic} is not a rigorous constraint, it provides a good indication of values of $u_{\mathrm{MV}}$ that can be chosen without affecting attainable accuracy and convergence. 

The computational complexity of FP-CG detection is significantly lower than that of CG detection due to the utilization of finite-precision arithmetic. For instance, in non-correlated ($\zeta = 0$) and moderately correlated channels ($\zeta = 0.5$), FP-CG detection can reduce computational cost by $75\%$ compared with CG detection (See Figs. \ref{fig: Heuristic_0} and \ref{fig: Heuristic_0p5}). In highly correlated channels ($\zeta=0.8$), FP-CG detection achieves a $43.3\%$ reduction in computational cost compared to CG detection (See Figs. \ref{fig: Heuristic_0p8} and \ref{fig: cc_con}).

\section{Joint Finite Precision and Preconditioning CG Detection}
\label{sec: joint}
In this section, we consider decreasing the iteration number based on FP-CG detection to further reduce the computational complexity of CG detection. First, the BJ preconditioning is introduced based on the correlated channel property, and then the FP-BJ-CG detection is proposed. Finally, we present the performance analysis of FP-BJ-CG detection.

\subsection{Block Jacobi Preconditioning and FP-BJ-CG Detection}
Our motivation stems from the observation that the Gram matrix of correlated channels in massive MIMO systems exhibits \textit{near block diagonal dominance}. This behavior is formalized in the following theorem.
\begin{theorem}[\textit{Asymptotic Analysis of Correlated MIMO Channel}]\label{the: correlated}
    Suppose the correlated channel matrix satisfying \eqref{eq: channel_model}, when $M$ is asymptotically large, i.e., $M\rightarrow \infty$, we have
    \begin{align}
        \mathbf{H}^{\mathsf{H}}\mathbf{H}\xrightarrow{a. s.}\mathbf{R}_t,
    \end{align}
    where $a.s.$ denotes almost sure convergence.
\end{theorem}
\begin{IEEEproof}
    The proof is available in Appendix \ref{app: correlated}.
\end{IEEEproof}
\textit{Theorem} \ref{the: correlated} shows that the Gram matrix of correlated channel asymptotically approaches deterministic exponential correlation matrices $\mathbf{R}_t$. Since  inter-user correlation is typically weak \cite{10653741}, the Gram matrix tends to be \textit{near block diagonally dominant} for massive MIMO systems\footnote{This behavior has been observed in \cite{bentrcia2019domain,9374779} by simulation. To our knowledge, this paper provides the first theoretical proof of this behavior.}, and the block property becomes more evident with the increase of correlation coefficient $\zeta$. Furthermore, when $\zeta = 0$, \textit{Theorem} \ref{the: correlated} can reduce to the well-known favorable propagation of the Rayleigh fading channel in \cite{6951994}.

Based on the near block diagonally dominant property, the BJ preconditioning \cite{eisenstat1981efficient} is introduced for joint finite precision and precondition CG detection (Appendix \ref{app: pre} compares the BJ preconditioning with other preconditioning methods). This can further reduce the computational complexity of CG detection by lowering the iteration count required for convergence. Specifically, the preconditioning matrix ${\bf P}_d$ uses the block diagonal entries of ${\bf A}$ and contains $d$ block sub-matrices. For simplification, the dimensions of block sub-matrices are assumed to be identical, denoted as $L=N/d\in \mathbb{Z}$ with $L<N$. Then the specific expression of ${\bf P}_d\in\mathbb{C}^{N\times N}$ and its inversion $\mathbf{M}_d\in\mathbb{C}^{N\times N}$ are given by
\begin{align}
    \mathbf{P}_d&=\mathrm{diag}\left( \mathbf{A}_1,\cdots ,\mathbf{A}_d \right),\\
    \mathbf{M}_d& = \mathbf{P}_d^{-1}=\mathrm{diag}\left( \mathbf{A}_{1}^{-1},\cdots ,\mathbf{A}_{d}^{-1} \right),\label{eq: pre_BJ}
\end{align}
where ${\bf A}_i \in \mathbb{C}^{L\times L}$ is the block diagonal part of ${\bf A}$ with $i=1,\cdots,d$.

\begin{algorithm}[t]
    \DontPrintSemicolon
    \SetAlgoNlRelativeSize{-1}
    \label{alg: FP-BJ-CG}
    \caption{FP-BJ-CG Detection with three precisions $u_{\rm MV}$, $u_{\rm IP}$ and $u_{\rm W}$}
    \KwIn{${\bf A}$, ${\bf b}$, the number of iteration $\mathcal{I}$, and dimension of each sub-matrix $L$}
    \KwOut{Estimated transmit signal vector $\hat{\bf x}$}

    Calculate ${\bf M}_d$ using \eqref{eq: pre_BJ} with $L$

    ${\bf r}_0 = {\bf b}$, ${\bf p}_0 = {\bf M}_d{\bf r}_0$, ${\bf x}_0 = {\bf 0}$

    \For{$i = 1:\mathcal{I}$}{

     \textcolor{red!60}{$\boldsymbol{ \varpi}_{i-1} = {\bf A}{\bf p}_{i-1}$ \Comment*[r]{$u_{\rm MV}$}} 

    ${\varphi}_{i-1} = {\bf p}_{i-1}^{\mathsf{H}}\boldsymbol{ \varpi}_{i-1}$ \Comment*[r]{$u_{\rm IP}$} 

    ${\varrho}_{i-1} = {{\bf r}_{i-1}^{\mathsf{H}}{\bf p}_{i-1}}$ \Comment*[r]{$u_{\rm IP}$}

     $\alpha_{i-1} ={\varrho}_{i-1} /{{\varphi}_{i-1}}$  \Comment*[r]{$u_{\rm W}$}

     ${\bf x}_i = {\bf x}_{i-1} + \alpha_{i-1}{\bf p}_{i-1} $ \Comment*[r]{$u_{\rm W}$}

    ${\bf r}_i = {\bf r}_{i-1} - \alpha_{i-1}\boldsymbol{ \varpi}_{i-1} $ \Comment*[r]{$u_{\rm W}$}

    ${\bf z} = {\bf M}_d{\bf r}_i $ \Comment*[r]{$u_{\rm W}$}

   ${\varrho}_{i} = {\bf r}_i^{\mathsf{H}}{\bf z}$ \Comment*[r]{$u_{\rm IP}$} 

    $\beta_{i} = {\varrho}_{i}/{{\varrho}_{i-1}}$ \Comment*[r]{$u_{\rm W}$}

    ${\bf p}_i = {\bf z} + \beta_{i}{\bf p}_{i-1}$ \Comment*[r]{$u_{\rm W}$}
    
    }

     $\hat{\bf x} = {\bf x}_{\mathcal{I}}$
    
\end{algorithm}

Given the BJ preconditioning, we present a basic implementation of FP-BJ-CG detection in \textit{Algorithm} \ref{alg: FP-BJ-CG}. More precisely, \textit{Algorithm} \ref{alg: FP-BJ-CG} employs a left-preconditioned CG method, which implicitly uses the asymmetric matrix ${\bf M}_d{\bf A}$ instead of the symmetric matrix $\Bar{\bf A}={\bf M}_d^{1/2}{\bf A}{\bf M}_d^{1/2}$. Both approaches produce identical sequences of iterates \cite{saad2003iterative}, yet the former is more computationally efficient. For convenience, the following theoretical analysis results are presented regarding the symmetrically preconditioned matrix, i.e., $\Bar{\bf A}$.

\subsection{Performance Analysis of FP-BJ-CG Detection}
This subsection presents the attainable accuracy, convergence, and computational complexity analysis of FP-BJ-CG detection. To simplify and clarify the analysis, we assume the $k$-th user has $N_{\rm UE}$ antennas, i.e., $N_k=N_{\rm UE}$ and $KN_{\rm UE}=N$.

First, the attainable accuracy of FP-BJ-CG detection in \textit{Algorithm} \ref{alg: FP-BJ-CG} is provided in the following theorem.
\begin{theorem}[\textit{Attainable Accuracy of \textit{Algorithm} \ref{alg: FP-BJ-CG}}]
\label{the: Attainable_BJ}
    The different with \textit{true} residuals $\Bar{{\bf b}} - \Bar{{\bf A}}\Bar{{\bf x}}_i$ and the \textit{computed} residuals vectors $\Bar{{\bf r}}_i$, i.e., attainable accuracy, satisfies
    \begin{align}
        \frac{\left\| \Bar{{\bf b}} - \Bar{{\bf A}}\Bar{{\bf x}}_i-\Bar{{\bf r}}_i \right\| _2}{\left\| \Bar{{\bf A}} \right\| _2\left\| \Bar{{\bf x}} \right\| _2}\le \left( 1+i \right) u_{\mathrm{W}}+ \varUpsilon_i\Theta _i, \label{eq: bj_mp}
    \end{align}
    where 
    \begin{align}
        \varUpsilon_i = &\left( \left( 8\sqrt{2}+6 \right) i+1 \right) u_{\mathrm{W}} \notag \\
        &+\left( 2i+1 \right) 2\sqrt{2}N^{\frac{3}{2}}u_{\mathrm{MV}}, \\
        \Theta _i&=\max_{k\le i} \frac{\left\| \Bar{{\bf x}}_k \right\| _2}{\left\| \Bar{{\bf x}} \right\| _2}.
    \end{align}
    Further, provided that \textit{Algorithm} \ref{alg: FP-BJ-CG} converges, the attainable accuracy satisfies
    \begin{align}
        \frac{\left\| \Bar{{\bf b}} - \Bar{{\bf A}}\Bar{{\bf x}}_i-\Bar{{\bf r}}_i \right\| _2}{\left\| \Bar{{\bf A}} \right\| _2\left\| \Bar{{\bf x}} \right\| _2}&\le \left( 1+i \right) u_{\mathrm{W}}+\varUpsilon_i \left( 1+\sqrt{\kappa \left( \Bar{{\bf A}} \right)} \right) \notag \\
        &\sim \mathcal{O} \left( N^{\frac{3}{2}}u_{\mathrm{MV}}\sqrt{\kappa \left( \Bar{{\bf A}} \right)} \right), \label{eq: bj_or_ph}
    \end{align}
    where 
    \begin{align}
        \Bar{\bf A}={\bf M}_d^{1/2}{\bf A}{\bf M}_d^{1/2},~\Bar{{\bf b}} = {\bf M}_d^{1/2} {\bf b},~\Bar{{\bf x}}={\bf M}_d^{-1/2}{\bf x}.
    \end{align}
\end{theorem}
\begin{IEEEproof}
    For a symmetric positive definite precondition ${\bf M}_d$ is used for FP-BJ-CG detection, it is equivalent to applying the un-preconditioning algorithm to the problem $\Bar{\bf A}\Bar{{\bf x}}=\Bar{{\bf b}}$, where $\Bar{\bf A}={\bf M}_d^{1/2}{\bf A}{\bf M}_d^{1/2},\Bar{{\bf b}} = {\bf M}_d^{1/2} {\bf b},\Bar{{\bf x}}={\bf M}_d^{-1/2}{\bf x}$. Then, similar to the proof of \textit{Theorem} \ref{the: Attainable}, we obtain \textit{Theorem} \ref{the: Attainable_BJ}.
\end{IEEEproof}
\textit{Theorem} \ref{the: Attainable_BJ} demonstrates that the attainable accuracy of FP-BJ-CG detection depends on work precision $u_{\rm W}$, matrix-vector products precision $u_{\rm MV}$, the dimension of $\Bar{\bf A}$, and condition number $\kappa \left( \Bar{\bf A} \right)$. Furthermore, similar to the analysis of \textit{Theorem} \ref{the: Attainable}, we can obtain the same results of \textit{Remarks} \ref{rem: impact_fp} and \ref{rem: con}. Moreover, the \textit{heuristic} method is changed into
\begin{align}
    u_{\mathrm{MV}}<\frac{1}{N^{1.5}\sqrt{\kappa \left( \Bar{\bf A} \right)}} \label{eq: heuristic_bj}.
\end{align}

Then, we compare the convergence of FP-CG detection and FP-BJ-CG detection via the condition numbers $\kappa\left(\Bar{\bf A}\right)$ and $\kappa\left({\bf A}\right)$. First, following the work of \cite{10653741, 9374779}, we make a common assumption for correlated channels in multi-user MIMO systems:
\begin{assumption}\label{ass: r}
    Let ${\bf R}_t^{k,j}$ be a block of ${\bf R}_t$ representing the correlation between the $k$-th and $j$-th user. Then, ${\bf R}_t^{k,j} = {\bf 0}$ for $\forall k\neq j$.
\end{assumption}
Using \textit{Assumption} \ref{ass: r} and \textit{Theorem} \ref{the: correlated}, we derive the following result:
\begin{align}
    \lim_{M\rightarrow \infty} {\bf H}_k^\mathsf{H}{\bf H}_j = 0, ~ \forall k\neq j. \label{eq: lim_M}
\end{align}
In massive MIMO systems, the number of BS antennas can reach hundreds or even thousands. Thus, the condition in \eqref{eq: lim_M} can be approximated by the following assumption:
\begin{assumption}\label{ass: h}
    Under \textit{Assumption} \ref{ass: r}, for massive MIMO systems, \eqref{eq: lim_M} can be approximated as follows:
    \begin{align}
        {\bf H}_k^\mathsf{H}{\bf H}_j = 0, ~ \forall k\neq j.
    \end{align}
\end{assumption}
\textit{Assumption} \ref{ass: h} is referred to spatial orthogonality \cite{8861014}. Using \textit{Assumption} \ref{ass: h}, we can compare the $\kappa\left(\Bar{\bf A}\right)$ and $\kappa\left({\bf A}\right)$, which is summarized in the following theorem.
\begin{theorem}[\textit{Impact of BJ Preconditioning on Convergence}]
    \label{the: compare_Condition}
    Under \textit{Assumption} \ref{ass: h}, given ${\bf M}_d$ in \eqref{eq: pre_BJ} and $N_{\rm UE}\leq L$, we have
    \begin{align}
         \kappa\left(\Bar{\bf A}\right) < \kappa\left({\bf A}\right).\label{eq: convergence}
    \end{align}
\end{theorem}
\begin{IEEEproof}
    The proof is available in Appendix \ref{app: compare_Condition}.
\end{IEEEproof}
\textit{Theorem} \ref{the: compare_Condition} indicates that FP-BJ-CG detection converges faster than FP-CG detection. Moreover, due to the lower condition number, FP-BJ-CG detection can support lower-precision arithmetic than FP-CG detection, according to \eqref{eq: heuristic} and \eqref{eq: heuristic_bj}. 

Finally, we present the computational complexity of FP-BJ-CG detection. The computational complexity of FP-BJ-CG detection comprises two components: preconditioning and the iterative process. For the preconditioning process, the computational complexity is mainly caused by the inversion of the block diagonal matrix ${\bf P}_d$, leading to $d\times \mathcal{O}\left(L^3\right)=\mathcal{O}\left(NL^2\right)$. In the iterative process, since ${\bf M}_d$ is block diagonal, the computational complexity of ${\bf M}_d{\bf r}_i$ in \textit{Step 10} is $\mathcal{O}\left(NL\right)$, still lower than that of ${\bf A}{\bf p}_{i-1}$ in \textit{Step 4}. Therefore, ${\bf A}{\bf p}_{i-1}$ dominant the computational complexity of the iterative process and requires $\mathcal{O}\left(N^2\right)$. In summary, the overall computational complexity of FP-BJ-CG detection in \textit{Algorithm} \ref{alg: FP-BJ-CG} is $\mathcal{O}\left(NL^2+\mathcal{I}N^2\right)$.  

\begin{remark}[\textit{Computational Complexity Reduction Compared with CG Detection}]
    Despite the extra computational complexity of BJ preconditioning, FP-BJ-CG detection can reduce the computational complexity by decreasing the iteration number and supporting lower-precision arithmetic. For example, considering $N_{\rm UE} =4$, $N=32$, and $L=8$, the extra computational complexity of BJ preconditioning is equivalent to $2$ CG detection iterations. Notably, Fig. \ref{fig: pcg} shows that FP-BJ-CG detection can accelerate CG detection by up to approximately $5$ iterations and support half-precision arithmetic ($\mathtt{fp16}$) to achieve the accuracy of LMMSE detection. Specifically, FP-BJ-CG detection with $\mathtt{fp16}$ decreases $30.0\%$ and $78.2\%$ computational cost compared to CG and LMMSE detection with $\mathtt{fp64}$, respectively (See Fig. \ref{fig: cc_10}). These advantages of FP-BJ-CG detection significantly offset its additional preconditioning cost and reduce the overall computational complexity.
\end{remark}

\begin{remark}[\textit{Trade-off Analysis between FP-CG and FP-BJ-CG Detection}]\label{rem: trade}
    The computational complexity of the preconditioning step is $\mathcal{O}\left(NL^2\right)$. Although this represents a one-time cost per channel realization, it may constitute a non-trivial overhead compared with FP-CG detection. To quantitatively assess this trade-off, we derive a general break-even criterion that identifies when preconditioning becomes computationally advantageous. Let $C_\mathrm{BJ}$ denote the preconditioning setup cost and let $C_\mathrm{PCG}^\mathrm{it}$ and $C_\mathrm{CG}^\mathrm{it}$ denote the per-iteration computational cost of the preconditioned and non-preconditioned FP-CG methods, respectively. Assuming FP-BJ-CG and FP-CG use the same finite-precision arithmetic, the total costs are
    \begin{align}
    C_\mathrm{PCG} = C_\mathrm{BJ} + k_\mathrm{PCG}C_\mathrm{PCG}^\mathrm{it},~C_\mathrm{CG} = k_\mathrm{CG}C_\mathrm{CG}^\mathrm{it},
    \end{align}
    where $k_\mathrm{PCG}$ and $k_\mathrm{CG}$ are the number of iterations by the preconditioned and non-preconditioned algorithms to reach the target accuracy. Notably, the preconditioning becomes computationally advantageous when $C_\mathrm{PCG}<C_\mathrm{CG}$, i.e.,
    \begin{align}
    k_\mathrm{CG}-k_\mathrm{PCG}>\frac{C_\mathrm{BJ}}{C_\mathrm{CG}^\mathrm{it}} + k_\mathrm{PCG}\left(\frac{C_\mathrm{PCG}^\mathrm{it}}{C_\mathrm{CG}^\mathrm{it}} - 1\right).
    \end{align}
    Since the dominant per-iteration cost is the same matrix–vector product for both methods, we have $C_\mathrm{PCG}^\mathrm{it}\approx C_\mathrm{CG}^\mathrm{it}\triangleq C^\mathrm{it}$. Under this assumption, the break-even condition simplifies to
    \begin{align}
        \Delta k \triangleq k_\mathrm{CG}-k_\mathrm{PCG}>\frac{C_\mathrm{BJ}}{C^\mathrm{it}}.
    \end{align}
    Substituting the complexity scalings $C_\mathrm{BJ}=\mathcal{O}\left(NL^2\right)$ and $C^\mathrm{it}=\mathcal{O}\left(N^2\right)$, we obtain
    \begin{align}\label{eq: break-even}
        \Delta k > \mathcal{O}\left(\frac{L^2}{N}\right).
    \end{align}
    From \eqref{eq: break-even}, for a certain user-side antennas $N$, the break-even number of iterations saved scales approximately with $L^2$. Equivalently, for a fixed block size $L$, the required iteration savings decrease as $N$ increases. \textit{This indicates that, for large $N$ and moderate block sizes $L$, the BJ preconditioning cost is easily amortized, and FP-BJ-CG is computationally advantageous. Conversely, for small $N$ and large $L$, the preconditioning overhead may outweigh the benefits, and FP-CG becomes preferable.} Simulation results in Sec. \ref{sec: sim} will provide the simulated break-even points.
\end{remark}

\begin{remark}[\textit{Case When \textit{Assumption} \ref{ass: r} and \ref{ass: h} Does Not Hold}]
When \textit{Assumption} \ref{ass: r} and \ref{ass: h} does not hold, we define $\mathbf{M}*$ as the optimal preconditioning matrix and have the following inequality \cite{doi:10.1137/0720040,bj_nr}:
\begin{align}\label{eq: in_ass} 
\kappa\left(\mathbf{M}_*^{1/2}{\bf A}\mathbf{M}_*^{1/2}\right)\leq \kappa\left(\Bar{\bf A}\right)\leq d\cdot\kappa\left(\mathbf{M}_*^{1/2}{\bf A}\mathbf{M}_*^{1/2}\right).
\end{align}
As shown in \eqref{eq: in_ass}, when \textit{Assumption} \ref{ass: r} and \ref{ass: h} is relaxed, the BJ preconditioning no longer yields the optimal condition number. Instead, its condition number lies within a bounded narrow interval $\left[ \kappa\left(\mathbf{M}_*^{1/2}{\bf A}\mathbf{M}_*^{1/2}\right), d\cdot\kappa\left(\mathbf{M}_*^{1/2}{\bf A}\mathbf{M}_*^{1/2}\right)\right]$. 
\end{remark}

\section{Simulation Results and Discussion}
\label{sec: sim}
In this section, we will provide numerical results to verify our derived results. First, the simulation setup is provided. Then, we validate the insights from \textit{Remarks} \ref{rem: impact_fp} and \ref{rem: con} and the heuristic method of FP-CG detection. Finally, the superiority of FP-BJ-CG detection is verified.

\subsection{Simulation Setup}
\subsubsection{Simulating Finite-Precision Arithmetic} The authors in \cite{higham2019simulating} provided a function that can be utilized to simulate $\mathtt{fp32}$, $\mathtt{fp16}$, and other low-precision arithmetic. For the analysis of computational cost, we consider a ratio of $1:2:4$ for the costs of $\mathtt{fp16}$, $\mathtt{fp32}$, and $\mathtt{fp64}$ arithmetic \cite{FAM}.

\subsubsection{Simulation Parameters} The BS antennas is set to be $M=256$ serving $K=8$ users. Each user is equipped with $N_{\rm UE}=4$ antennas. Thus, $N=N_{\rm UE}K=32$. We use the correlated channel model in \eqref{eq: channel_model}. 16QAM modulation scheme is adopted. For the BJ preconditioning, the dimension of block sub-matrices is $L=8$. Moreover, we set $500,000$ Monte-Carlo trials for each iteration or SNR level.

\begin{figure*}[t]
    \centering
    \subfloat[$\zeta = 0$]{\includegraphics[width=0.32\textwidth]{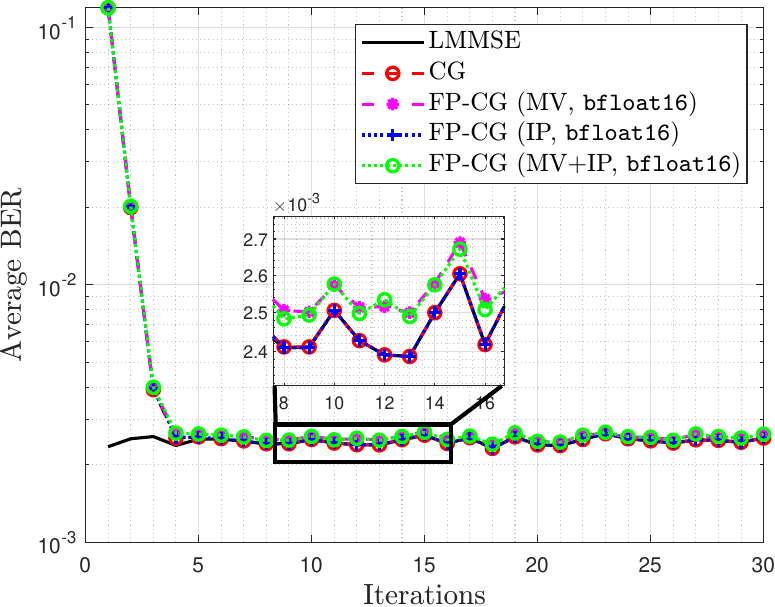}
    \label{fig: ip_mv_0}}
    \subfloat[$\zeta = 0.5$]{\includegraphics[width=0.32\textwidth]{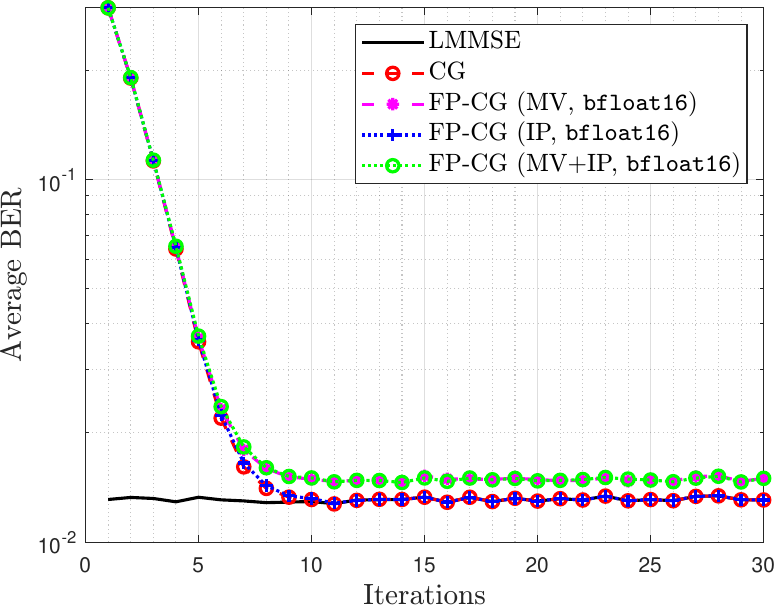}
    \label{fig: ip_mv_0p5}}
    \subfloat[$\zeta = 0.8$]{\includegraphics[width=0.32\textwidth]{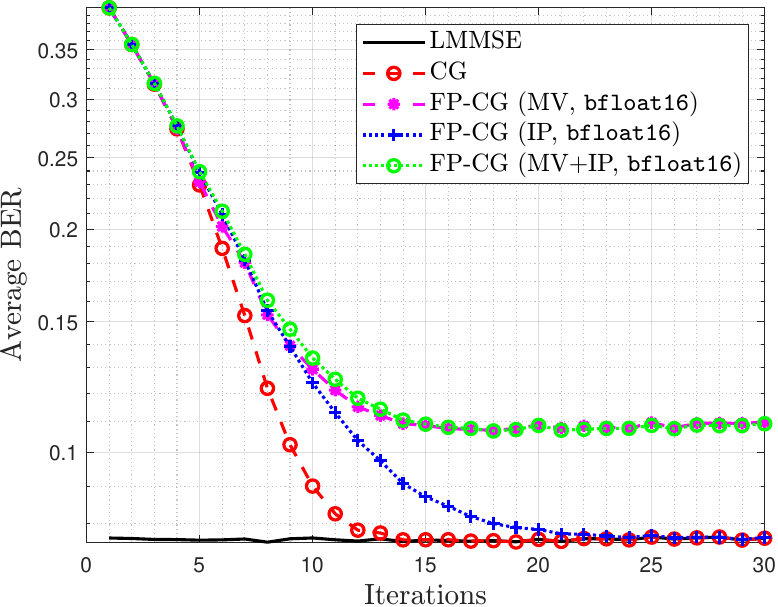}
    \label{fig: ip_mv_0p8}}
    \caption{Convergence curve of FP-CG detection using $\mathtt{bfloat16}$ with different $\zeta$ in SNR $=20$ dB. MV: matrix-vector products using $\mathtt{bfloat16}$. IP: inner products using $\mathtt{bfloat16}$. MV + IP: both using $\mathtt{bfloat16}$.}
    \label{fig: ip_mv}
\end{figure*}

\begin{figure*}[t]
    \centering
    \subfloat[$\zeta = 0$]{\includegraphics[width=0.32\textwidth]{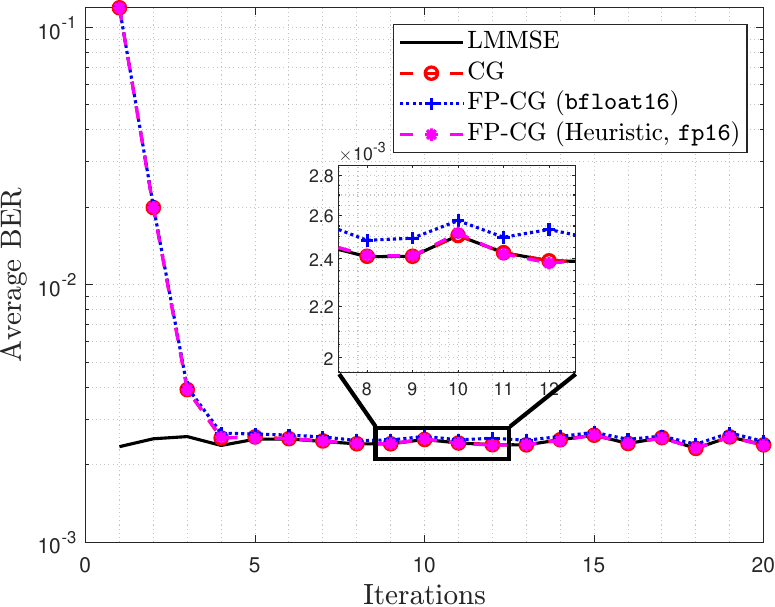}
    \label{fig: Heuristic_0}}
    \subfloat[$\zeta = 0.5$]{\includegraphics[width=0.32\textwidth]{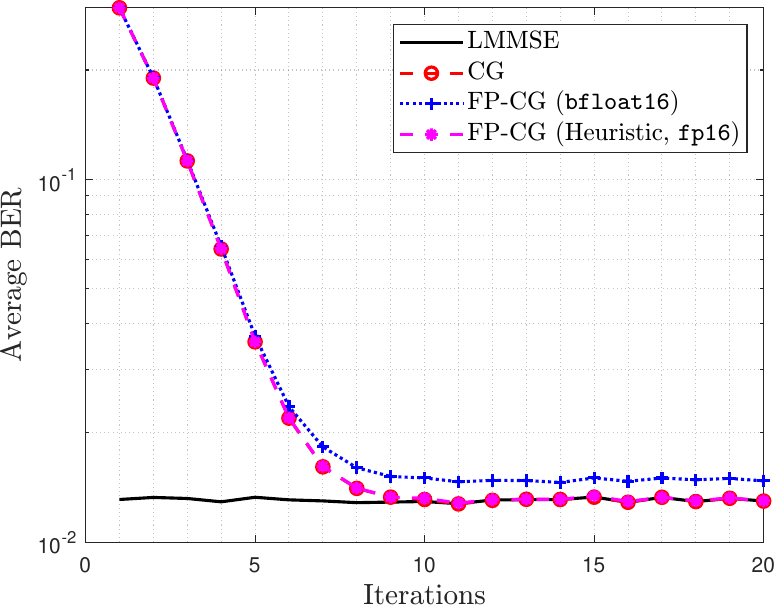}
    \label{fig: Heuristic_0p5}}
    \subfloat[$\zeta = 0.8$]{\includegraphics[width=0.32\textwidth]{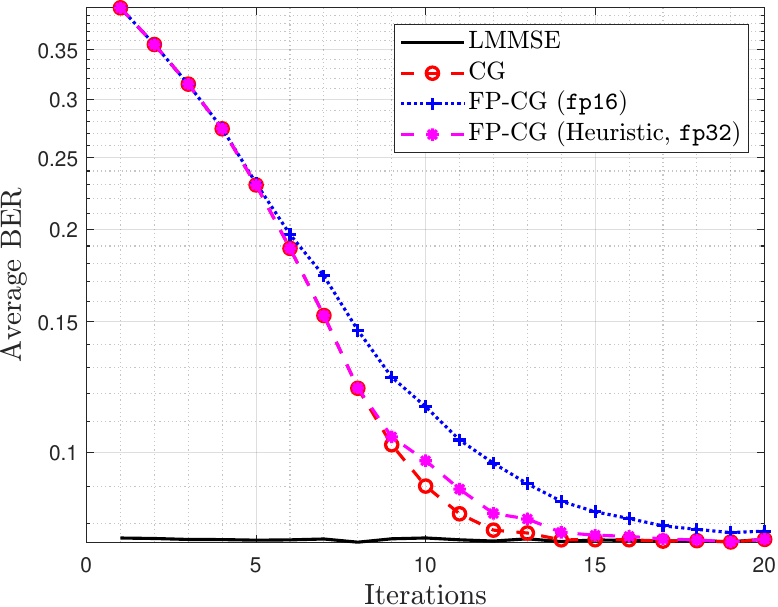}
    \label{fig: Heuristic_0p8}}
    \caption{Convergence curve of FP-CG detection using the heuristic method with different $\zeta$ in SNR $=20$ dB.}
    \label{fig: Heuristic}
\end{figure*}

\subsection{Performance of FP-CG Detection}
\subsubsection{Impact of Finite-Precision Arithmetic} As depicted in Fig. \ref{fig: ip_mv}, we compare FP-CG detection in three cases: when only matrix-vector products use $\mathtt{bfloat16}$, only inner products use $\mathtt{bfloat16}$, and both use $\mathtt{bfloat16}$, under varying channel correlation in SNR $=20$ dB. The results indicate that using low-precision arithmetic for inner products does not impact the attainable accuracy of FP-CG detection, whereas matrix-vector products have a significant effect on the attainable accuracy. Furthermore, as channel correlation increases, i.e., as the channel condition number grows, the gap and convergence delay between CG and FP-CG detection become more pronounced. These observations are consistent with the insights from \textit{Remarks} \ref{rem: impact_fp} and \ref{rem: con}.

\subsubsection{Heuristic Method} Fig. \ref{fig: Heuristic} illustrates the convergence curve of FP-CG detection employing the heuristic method for different correlation coefficients $\zeta$ in SNR $=20$ dB. For example, when $\zeta = 0.5$, we first compute $u_{\rm MV}<1.33\times 10^{-3}$ using \eqref{eq: heuristic}, then refer to Table \ref{tab: parafp}, and ultimately select $\mathtt{fp16}$. As shown in Fig. \ref{fig: Heuristic_0p5}, the heuristic method ($\mathtt{fp16}$) achieves the same accuracy as CG detection, while the use of $\mathtt{bfloat16}$ leads to a noticeable performance loss. Similar trends are observed in Figs.~\ref{fig: Heuristic_0} and \ref{fig: Heuristic_0p8}. Moreover, Fig. \ref{fig: Heuristic} clearly shows that the heuristic method provides a practical guideline for selecting precision to ensure FP-CG detection convergence. Overall, the heuristic method serves as a valuable guideline for achieving attainable accuracy and ensuring convergence. Further, the simulation results confirm that the bound is tight, as performance degradation becomes evident when the precision does not hold the predicted threshold (the heuristic method).

\subsection{Performance of FP-BJ-CG Detection}
\subsubsection{Convergence Analysis} 
\begin{figure}[t]
     \centering
     \includegraphics[width=0.4\textwidth]{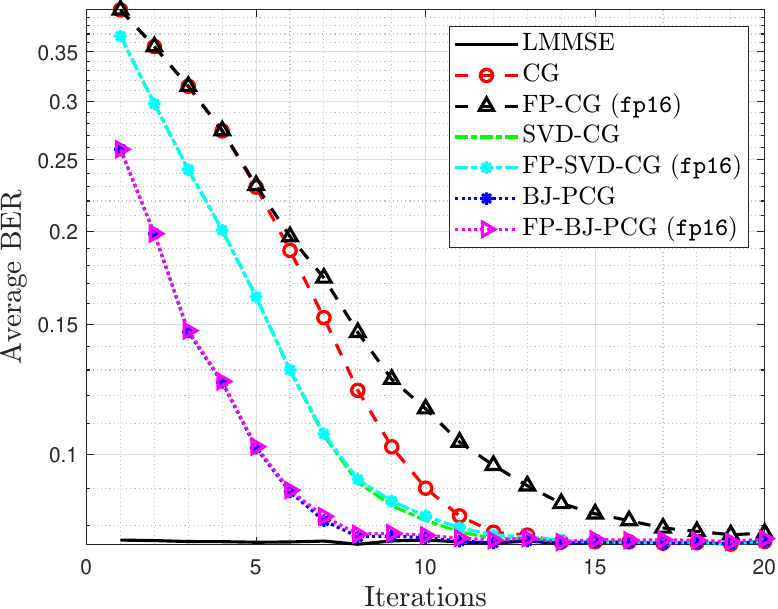}
     \caption{Convergence curve of various detection using $\mathtt{fp16}$ with $\zeta = 0.8$ in SNR $=20$ dB.}
     \label{fig: pcg}
 \end{figure}

In Fig. \ref{fig: pcg}, we compare the convergence curve of FP-BJ-CG Detection with different detection with $\zeta = 0.8$ in SNR $=20$ dB, where SVD-CG detection applies the pre-processing method from \cite{10653741} to CG detection. It can be observed that FP-BJ-CG detection achieves the fastest convergence among the compared methods. Since FP-BJ-CG detection has a lower condition number (See \textit{Theorem} \ref{the: compare_Condition}), $\mathtt{fp16}$ can be used for FP-BJ-CG detection based on the heuristic method instead of $\mathtt{fp32}$ for FP-CG detection, as shown in Fig. \ref{fig: Heuristic_0p8}.

\subsubsection{BER Performance} 
\begin{figure}[t]
     \centering
     \includegraphics[width=0.4\textwidth]{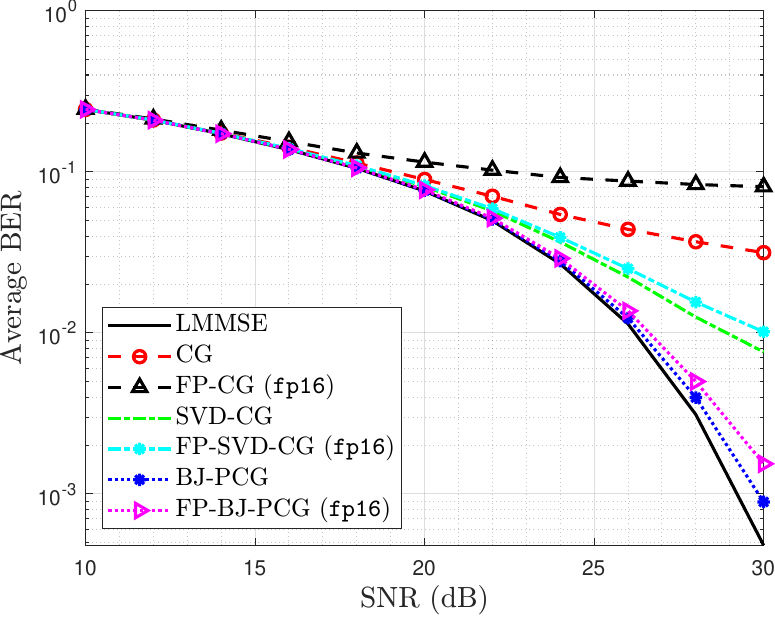}
     \caption{BER performance of various detection against SNR with $\zeta = 0.8$ and $\mathcal{I} = 10$.}
     \label{fig: ber_pcg}
 \end{figure}

As shown in Fig. \ref{fig: ber_pcg}, we compare the BER performance of FP-BJ-CG detection with different detection with $\zeta = 0.8$ and $\mathcal{I} = 10$. It can be observed that FP-BJ-CG detection with $\mathtt{fp16}$ exhibits similar performance to LMMSE detection at low SNR levels, with only an almost $1.2$ dB performance loss at high SNR levels. 

 \begin{figure}[t]
    \centering
    \subfloat[$\varepsilon = 20$ dB]{\includegraphics[width=0.4\textwidth]{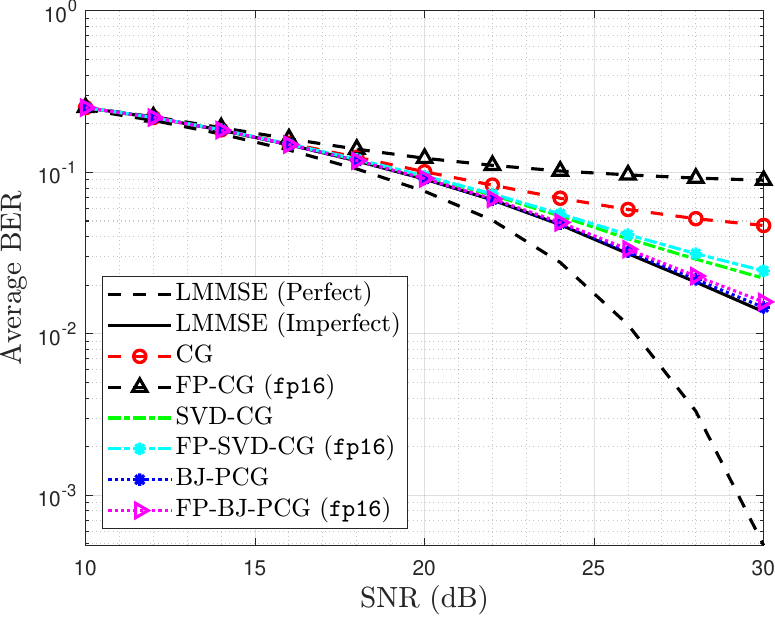}
    \label{fig: ber_pcg_im_20}} \\
    \subfloat[$\varepsilon = 25$ dB]{\includegraphics[width=0.4\textwidth]{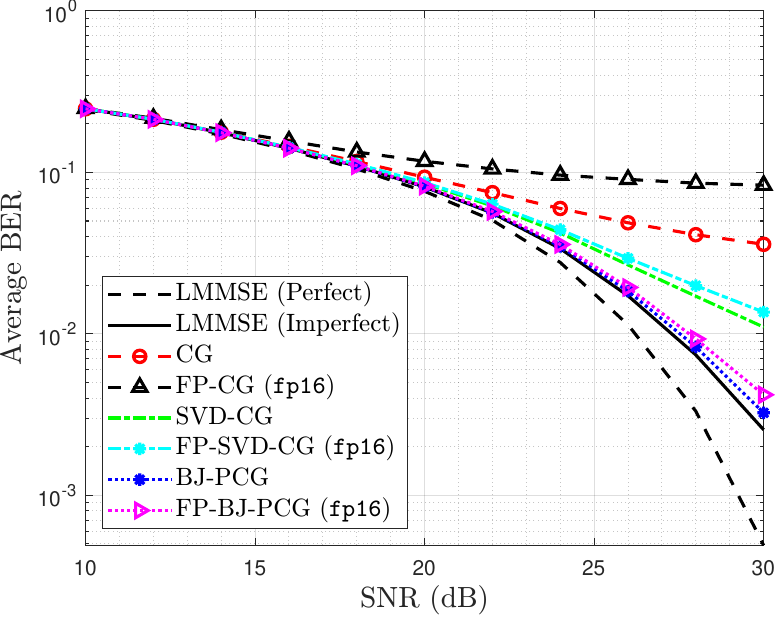}
    \label{fig: ber_pcg_im_25}}
    \caption{BER performance of various detection against SNR with imperfect CSI, $\zeta = 0.8$ and $\mathcal{I} = 10$.}
    \label{fig: ber_pcg_im}
\end{figure}

 \begin{figure}[t]
    \centering
    \subfloat[{Geometry-based channel.}]{\includegraphics[width=0.4\textwidth]{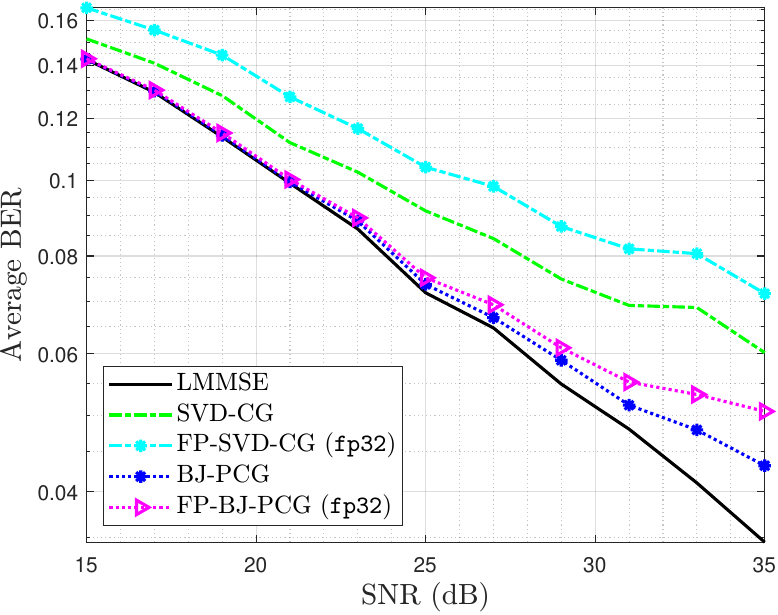}
    \label{fig: geo}} \\
    \subfloat[{Time-varying channel.}]{\includegraphics[width=0.4\textwidth]{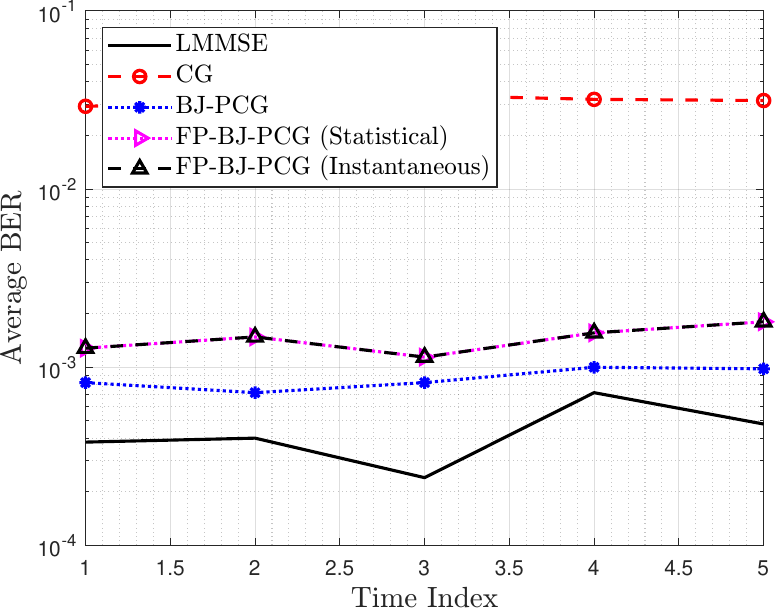}
    \label{fig: ber_time}}
    \caption{{BER performance of various detection against SNR with different channel conditions.}}
    \label{fig: channel}
\end{figure}

   \begin{figure}[t]
     \centering
     \includegraphics[width=0.4\textwidth]{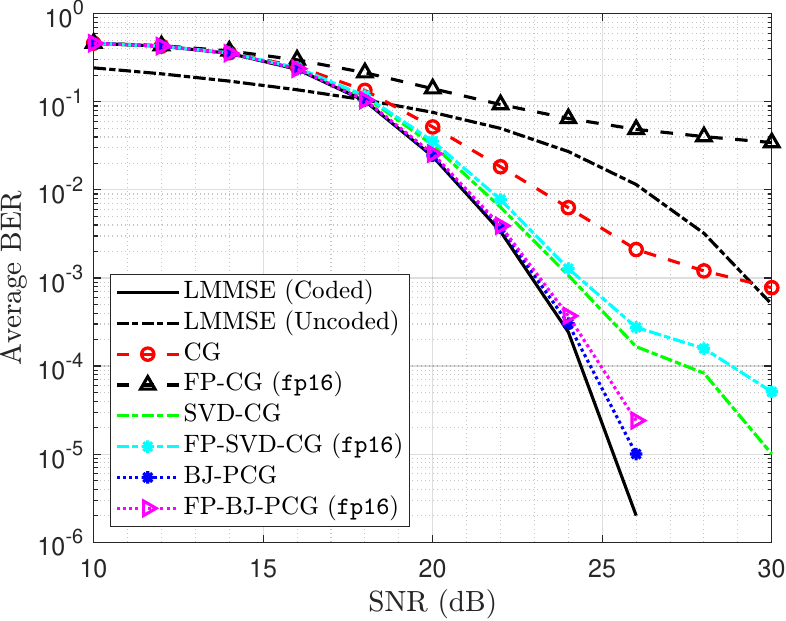}
     \caption{BER performance of various detection against SNR with convolutional coding, $\zeta = 0.8$ and $\mathcal{I} = 10$.}
     \label{fig: coding}
 \end{figure}

   \begin{figure}[t]
    \centering
    \subfloat[Various detection until convergence with $\zeta = 0.8$ in SNR $=20$ dB, where FP-CG detection with $\mathtt{fp32}$ and FP-BJ-CG detection with $\mathtt{fp16}$.]{\includegraphics[width=0.4\textwidth]{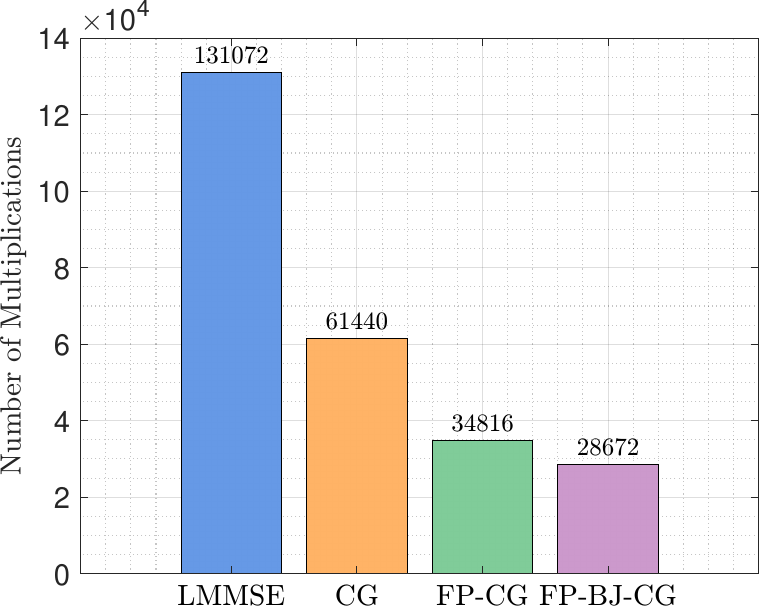}
    \label{fig: cc_con}} \\
    \subfloat[Various detection with $\mathcal{I} = 10$, where both FP-CG and FP-BJ-CG detection with $\mathtt{fp16}$.]{\includegraphics[width=0.4\textwidth]{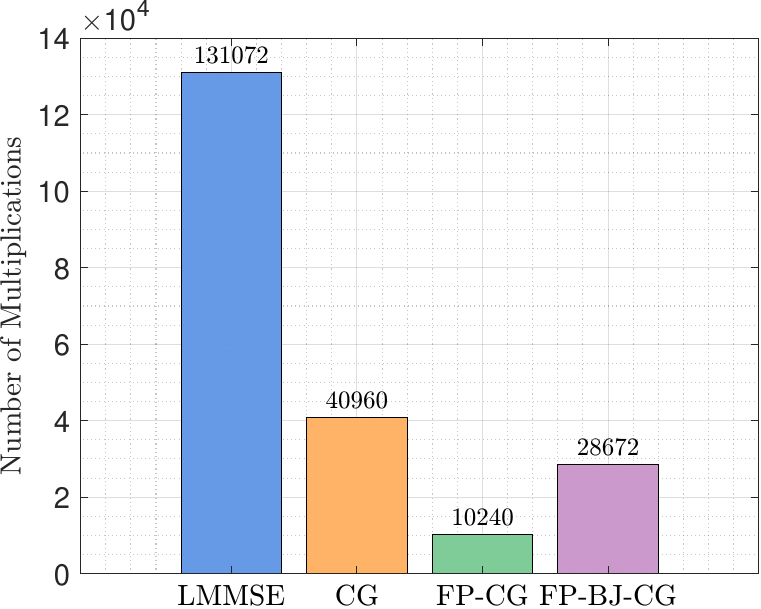}
    \label{fig: cc_10}}
    \caption{Number of multiplications for various detection.}
    \label{fig: cc}
\end{figure}

To further evaluate the robustness of FP-BJ-CG detection, imperfect channel state information (CSI) is considered in Fig. \ref{fig: ber_pcg_im}. Specifically, the imperfect CSI matrix can be expressed as \cite{10103510,10368038}
\begin{align}
    \Hat{\bf H} = {\bf H} + \Delta {\bf H},
\end{align}
where $\Delta {\bf H}\in\mathbb{C}^{M\times N}$ is the noise matrix. The ratio $\varepsilon$ between the power of the channel and noise elements is set to be $20$ dB and $25$ dB. Since LMMSE detection under imperfect CSI achieves only sub-optimal performance, FP-BJ-CG detection also converges to a sub-optimal solution. Nevertheless, FP-BJ-CG detection achieves performance comparable to LMMSE detection under imperfect CSI, demonstrating its robustness.

{ 
Moreover, the robustness of the proposed method under geometry-based channel models and time-varying conditions is shown in Fig. \ref{fig: channel}. Specifically, for the geometry-based channel, we have
\begin{equation}
{\bf H}_{k} = \sqrt{\frac{MN_{k}}{L_k}}\left(\alpha_{k,0}{\bf a}_t(\phi_{k,0}^{t}){\bf a}_r^\mathsf{H}(\phi_{k,0}^{r}) + \sum_{i=1}^{L_k-1}\alpha_{k,i}{\bf a}_t(\phi_{k,i}^{t}){\bf a}_r^\mathsf{H}(\phi_{k,i}^{r})\right),
\end{equation}
where $L_k$ denotes the number of propagation paths. The first term corresponds to the line-of-sight (LoS) component with $\alpha_{k,0}\sim \mathcal{CN}(0,1)$, while the remaining terms represent non-line-of-sight (NLoS) components with $\alpha_{k,i}\sim \mathcal{CN}(0,10^{-\mu})$. The angles of departure (AoD) $\phi_{k,i}^{t}$ and angles of arrival (AoA) $\phi_{k,i}^{r}$ are independently drawn from a uniform distribution $\mathcal{U}(-\pi,\pi)$. Both the base station and users are equipped with uniform linear arrays (ULAs) with half-wavelength spacing, and the array response vector is given by
\begin{equation}
{\bf a}(\phi)=\frac{1}{\sqrt{N}}[1,e^{j2\pi \frac{d}{\lambda} \sin{\phi}}, \cdots, e^{j2\pi \frac{d}{\lambda}(N-1) \sin{\phi}}]^\mathsf{T}.\notag
\end{equation}
We consider $L_k = 5$ and $\mu=1$, and the carrier frequency is $28 {\rm GHz}$. Other parameters are the same as those in the main paper. As shown in Fig.~\ref{fig: geo}, the proposed preconditioning method still achieves performance close to the LMMSE detection and consistently outperforms the SVD-based preconditioning. These results demonstrate that the proposed method remains effective under realistic multi-path fading conditions with varying propagation characteristics.

For the time-varying channel, we adopt the commonly used first-order Gauss-Markov model, where the channel evolves as \cite{7895158}
\begin{equation}
\mathbf{H}_i = \rho_\tau \mathbf{H}_{i-1} + \sqrt{1-\rho_\tau^2}\,\mathbf{W}_i,
\end{equation}
where $i$ is the time index, $\rho_\tau = J_0(2\pi f_d \tau)$, and $\mathbf{W}_i$ has the same spatial covariance as $\mathbf{H}_{i-1}$. Fig.~\ref{fig: ber_time} demonstrates that the BER performance of the heuristic is the same as the ideal case where instantaneous optimal precision is used, even under high Doppler ($f_d=1$ kHz) and strong spatial correlation ($\zeta = 0.8$). As a result, the heuristic does not require re-estimation at every coherence interval.
}

In addition, Fig. \ref{fig: coding} shows the BER performance of various detection methods against SNR under channel coding. We consider convolutional coding with a code rate of $\frac{1}{2}$ and employ a Viterbi decoder. The modulation scheme is 16-QAM. As illustrated in Fig. \ref{fig: coding}, the FP-BJ-CG detector using $\mathtt{fp16}$ precision achieves performance comparable to that of the LMMSE detector, demonstrating that the proposed FP-BJ-CG detection method performs effectively in coded massive MIMO systems.

\subsubsection{Computational Complexity Analysis} As demonstrated in Figs. \ref{fig: Heuristic_0p8} and \ref{fig: pcg}, achieving the accuracy of LMMSE detection requires $15$, $17$, and $10$ iterations for CG, FP-CG, and FP-BJ-CG detection, respectively. Based on this, Fig. \ref{fig: cc_con} compares the computational complexity of LMMSE, CG, FP-CG, and FP-BJ-CG detection until convergence under the same conditions as Fig. \ref{fig: pcg}. Specifically, FP-CG detection with $\mathtt{fp32}$ reduces computational overhead by $43.3\%$ and $73.4\%$ compared to CG and LMMSE detection with $\mathtt{fp64}$, respectively. Similarly, FP-BJ-CG detection with $\mathtt{fp16}$ decreases $53.3\%$ and $78.2\%$ computational consumption compared to CG and LMMSE detection with $\mathtt{fp64}$, respectively. {The computational complexity comparison at a high SNR level is shown in Appendix \ref{app: high_snr}.} Further, similar to the analysis in \textit{Example} \ref{ex: 1}, the required computational cost for FP-BJ-CG detection is on the order of $2.7$ TFLOPS. Therefore, FP-BJ-CG detection requires only about $27.9\%$ of the computational resources of the NVIDIA A100, indicating that it can be efficiently implemented on a modern GPU.

Moreover, Fig. \ref{fig: cc_10} shows the computational complexity of LMMSE, CG, FP-CG, and FP-BJ-CG detection with $10$ iterations, where FP-CG and FP-BJ-CG detection use $\mathtt{fp16}$. It can be observed that FP-BJ-CG detection with $\mathtt{fp16}$ decreases $30.0\%$ and $78.2\%$ computational cost compared to CG and LMMSE detection with $\mathtt{fp64}$, respectively. Although the computational complexity of FP-BJ-CG detection is a little higher than that of FP-CG detection, FP-BJ-CG detection achieves a near-BER performance than LMMSE detection, as shown in Figs. \ref{fig: ber_pcg} and \ref{fig: ber_pcg_im}.

 \begin{figure}[t]
    \centering
    \subfloat[Computational complexity of FP-CG and FP-BJ-CG as a function of $L$ with $N=32$ and $\zeta = 0.8$ in SNR $=20$ dB.]{\includegraphics[width=0.4\textwidth]{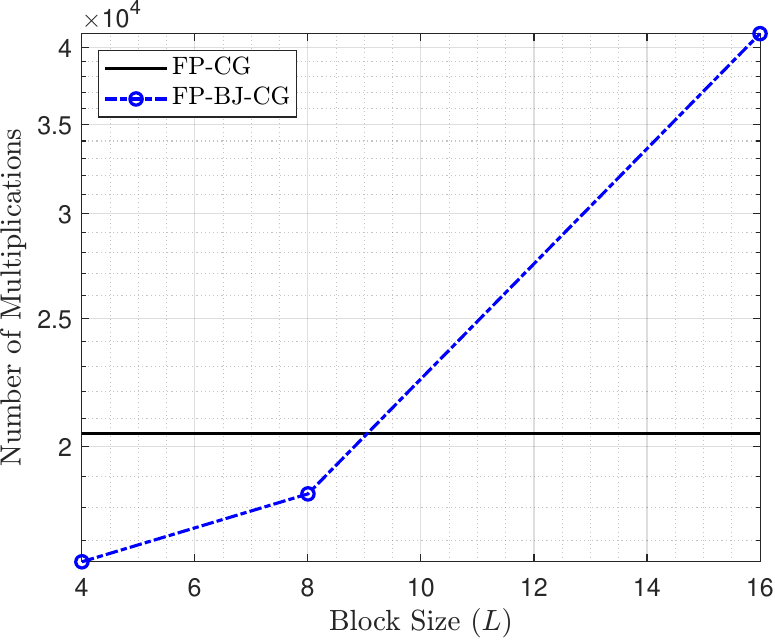}
    \label{fig: break_even_L}} \\
    \subfloat[Computational complexity of FP-CG and FP-BJ-CG as a function of $N$ with $L = 8$ and $\zeta = 0.8$ in SNR $=20$ dB.]{\includegraphics[width=0.4\textwidth]{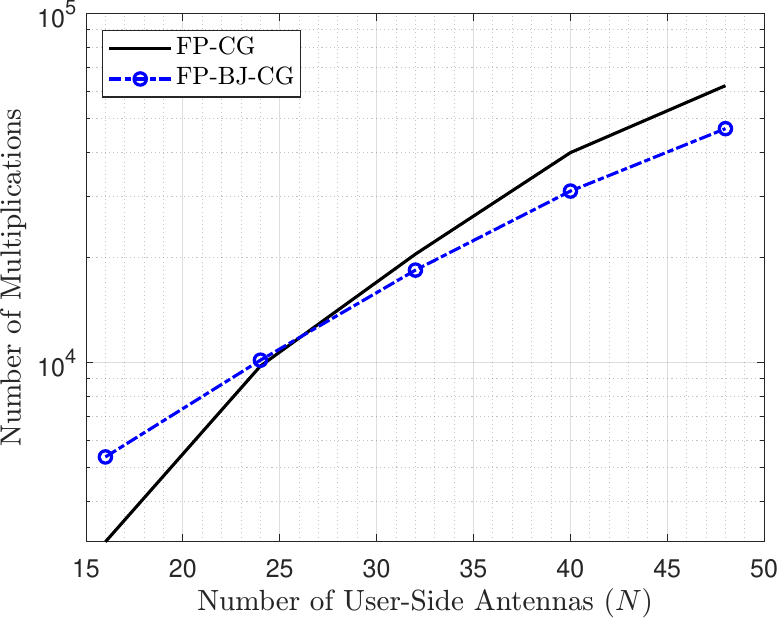}
    \label{fig: break_even_N}}
    \caption{Trade-off between FP-CG and FP-BJ-CG detection.}
    \label{fig: trade}
\end{figure}

Further, Fig. \ref{fig: trade} shows the simulated break-even points to further illustrate this trade-off between FP-CG and FP-BJ-CG detection. As depicted in Fig.~\ref{fig: break_even_L}, for $N=32$, the break-even point occurs at $L=9$, indicating that FP-BJ-CG is more efficient for $L \leq 9$, while FP-CG performs better for $L > 9$. Similarly, Fig.~\ref{fig: break_even_N} shows that for $L=8$, the break-even point is $N=26$. Thus, for $N \geq 26$, FP-BJ-CG outperforms FP-CG, consistent with the theoretical insight from \textit{Remark} \ref{rem: trade}.

{Overall, the computational savings of the proposed BJ‑FP‑CG method can directly translate into reduced hardware cost and power consumption. Each saved iteration eliminates a fixed number of multiply‑accumulate operations, thereby lowering dynamic power and enabling either higher throughput or lower clock frequency. Similar benefits can be observed for low‑precision arithmetic. From a system integration perspective, the method is an enhanced version of existing CG detection, requiring only the extraction of block diagonals. Moreover, many existing hardware platforms already support mixed‑precision arithmetic. Hence, the proposed approach is readily adoptable in current and next‑generation receivers.}

\section{Conclusions}
\label{sec: con}
In this paper, we have proposed a low computational complexity CG detection from a finite-precision perspective. First, we have presented FP-CG detection to alleviate the computational bottleneck of each CG iteration and analyzed the corresponding performance to provide insights into the impact of finite-precision arithmetic. Then, a practical heuristic has been developed, which can choose suitable precisions to avoid significant degradation in FP-CG detection accuracy under different channel conditions. Furthermore, we have proposed FP-BJ-CG detection to simultaneously reduce the number of iterations and the computational complexity per iteration, and the corresponding analyses of convergence and computational costs have been provided. Finally, simulation results have demonstrated that FP-BJ-CG detection achieves similar BER performance to LMMSE detection at low SNR levels, with only almost $1.2$ dB performance loss at high SNR levels. Considering computation complexity, FP-BJ-CG detection decreases $30.0\%$ and $78.2\%$ of the computational cost compared to CG and LMMSE detection, respectively.

\appendices
\section{Proof of \textit{Theorem} \ref{the: Attainable}}
\label{app: attainable_accuracy}
For \textit{Algorithm} \ref{alg: FP-CG}, \eqref{eq: update_of_x} and \eqref{eq: update_of_r} is computed in finite-precision arithmetic and satisfy
\begin{align}
    \mathbf{x}_i&=\mathbf{x}_{i-1}+\alpha _{i-1}\mathbf{p}_{i-1}+\boldsymbol{\xi }_i,\label{eq: fp_up_x}\\
    \mathbf{r}_i&=\mathbf{r}_{i-1}-\alpha _{i-1}\mathbf{Ap}_{i-1}+\boldsymbol{\eta }_i, \label{eq: fp_up_r}
\end{align}
where 
\begin{align}
    \left\| \boldsymbol{\xi }_i \right\| _2&\le u_{\mathrm{W}}\left\| \mathbf{x}_{i-1} \right\| _2+\tau _{\mathrm{W}}\left\| \alpha _{i-1}\mathbf{p}_{i-1} \right\| _2 \label{eq: xi}\\
    \left\| \boldsymbol{\eta }_i \right\| _2&\le u_{\mathrm{W}}\left\| \mathbf{r}_{i-1} \right\| _2+\tau _{\mathrm{W}}\left\| \alpha _{i-1}\mathbf{Ap}_{i-1} \right\| _2 \notag\\
    &+\upsilon _{\mathrm{W},\mathrm{MV}}\left\| \mathbf{A} \right\| _2\left\| \alpha _{i-1}\mathbf{p}_{i-1} \right\| _2, \label{eq: eta}\\
    \tau _{\mathrm{W}}&=u_{\mathrm{W}}+\sqrt{2}\gamma _{2}^{\left( \mathrm{W} \right)}+u_{\rm W}\sqrt{2}\gamma _{2}^{\left( \mathrm{W} \right)},\\
    \upsilon _{\mathrm{W},\mathrm{MV}}&=\left( u_{\rm W}+1 \right) \left( u_{\rm W}\sqrt{2}\gamma _{2}^{\left( \mathrm{W} \right)}+1 \right) \sqrt{2N}\gamma _{2N}^{\left( \mathrm{MV} \right)}.
\end{align}

Then, multiplying \eqref{eq: fp_up_r} with matrix $\bf A$ and subtracting from ${\bf b}$ obtains the \textit{true} residual ${\bf b-Ax}_{i}$. Further, subtracting from ${\bf r}_i$ in \eqref{eq: fp_up_r}, we have
\begin{align}
    \mathbf{b}-\mathbf{Ax}_i-\mathbf{r}_i&=\left( \mathbf{b}-\mathbf{Ax}_{i-1}-\mathbf{r}_{i-1} \right) -\mathbf{A}\boldsymbol{\xi }_i-\boldsymbol{\eta }_i\\
    &=\left( \mathbf{b}-\mathbf{Ax}_0-\mathbf{r}_0 \right) -\sum_{j=1}^i{\left( \mathbf{A}\boldsymbol{\xi }_j-\boldsymbol{\eta }_j \right)}.
\end{align}
Taking norms on both sides and dividing by $\left\| \mathbf{A} \right\| _2\left\| \mathbf{x} \right\| _2$ gives
\begin{align}
    \frac{\left\| \mathbf{b}-\mathbf{Ax}_i-\mathbf{r}_i \right\| _2}{\left\| \mathbf{A} \right\| _2\left\| \mathbf{x} \right\| _2}&\le \frac{\left\| \mathbf{b}-\mathbf{Ax}_0-\mathbf{r}_0 \right\| _2}{\left\| \mathbf{A} \right\| _2\left\| \mathbf{x} \right\| _2} \notag\\
    &+\sum_{j=1}^i{\left( \frac{\left\| \boldsymbol{\xi }_j \right\| _2}{\left\| \mathbf{x} \right\| _2}+\frac{\left\| \boldsymbol{\eta }_j \right\| _2}{\left\| \mathbf{A} \right\| _2\left\| \mathbf{x} \right\| _2} \right)}. \label{eq: b-ax}
\end{align}

Next, we determine the specific bounds of the right three terms in \eqref{eq: b-ax}. As the initial vector ${\bf r}_0$ can be obtained directly, the right first term in \eqref{eq: b-ax} can be derived based on \textit{Theorem} \ref{the: cv} and have
\begin{align}
    \left\| \mathbf{b}-\mathbf{Ax}_0-\mathbf{r}_0 \right\| _2&\le \left( 1+u_{\mathrm{W}} \right) \sqrt{2N}\gamma _{2N}^{\left( \mathrm{MV} \right)}\left\| \mathbf{A} \right\| _2\left\| \mathbf{x}_0 \right\| _2 \notag \\
    &+u_{\mathrm{W}}\left\| \mathbf{b} \right\| _2+u_{\mathrm{W}}\left\| \mathbf{A} \right\| _2\left\| \mathbf{x}_0 \right\| _2, \label{eq: all}
\end{align}
and note that $\left\| \mathbf{b} \right\| _2\le \left\| \mathbf{A} \right\| _2\left\| \mathbf{x} \right\| _2$, we have
\begin{align}
    \frac{\left\| \mathbf{b}-\mathbf{Ax}_0-\mathbf{r}_0 \right\| _2}{\left\| \mathbf{A} \right\| _2\left\| \mathbf{x} \right\| _2}&\le \left( 1+u_{\mathrm{W}} \right) \sqrt{2N}\gamma _{2N}^{\left( \mathrm{MV} \right)}\frac{\left\| \mathbf{x}_0 \right\| _2}{\left\| \mathbf{x} \right\| _2}\notag \\
    &+u_{\mathrm{W}}+u_{\mathrm{W}}\frac{\left\| \mathbf{x}_0 \right\| _2}{\left\| \mathbf{x} \right\| _2}. \label{eq: first_term}
\end{align}

Then, the right second term in \eqref{eq: b-ax} can be determined as follows. Using \eqref{eq: fp_up_x}, we can obtain
\begin{align}
    \alpha _{i-1}\mathbf{p}_{i-1}=\mathbf{x}_i-\mathbf{x}_{i-1}-\boldsymbol{\xi }_i, \label{eq: temp_x}
\end{align}
and substituting \eqref{eq: temp_x} into \eqref{eq: xi} gives
\begin{align}
    \left\| \boldsymbol{\xi }_i \right\| _2\le u_{\rm W}\left\| \mathbf{x}_{i-1} \right\| _2+\tau _{\mathrm{W}}\left( \left\| \mathbf{x}_i \right\| _2+\left\| \mathbf{x}_{i-1} \right\| _2+\left\| \boldsymbol{\xi }_i \right\| _2 \right).  \label{eq: temp_x1}
\end{align}
Note that $1-\tau_{\rm W} >0$ for using $\mathtt{fp64}$, \eqref{eq: temp_x1} can be expressed as
\begin{align}
    \left\| \boldsymbol{\xi }_i \right\| _2&\le \left( 2u_{\rm W}+\sqrt{2}\gamma _{2}^{\left( \rm W \right)} \right) \left\| \mathbf{x}_{i-1} \right\| _2+\left( u_{\rm W}+\sqrt{2}\gamma _{2}^{\left( \rm W \right)} \right) \left\| \mathbf{x}_i \right\| _2 \notag \\
    &+\mathcal{O} \left( u_{\rm W}^{2} \right) \left( \left\| \mathbf{x}_i \right\| _2+\left\| \mathbf{x}_{i-1} \right\| _2 \right), 
\end{align}
Thus, we have
\begin{align}
    \sum_{j=1}^i{\frac{\left\| \boldsymbol{\xi }_j \right\| _2}{\left\| \mathbf{x} \right\| _2}}\le \left( 3u_{\rm W}+2\sqrt{2}\gamma _{2}^{\left( \rm W \right)}+\mathcal{O} \left( u_{\rm W}^{2} \right) \right) i\Theta _i, \label{eq: second_term}
\end{align}
where 
\begin{align}
    \Theta _i=\max_{j\le i} \frac{\left\| \mathbf{x}_i \right\| _2}{\left\| \mathbf{x} \right\| _2}.
\end{align}

Moreover, we determine the right third term in \eqref{eq: b-ax}. Using \eqref{eq: temp_x}, the third term in \eqref{eq: eta} can be bounded by
\begin{align}
    \upsilon _{\mathrm{W},\mathrm{MV}}\left\| \mathbf{A} \right\| _2\left\| \alpha _{i-1}\mathbf{p}_{i-1} \right\| _2\le \sqrt{2N}\gamma _{2N}^{\left( \mathrm{MV} \right)}\left( 1+\mathcal{O} \left( u_{\rm W}^{2} \right) \right) \varXi _i \label{eq: eta3}
\end{align}
where 
\begin{align}
    \varXi _i=\left\| \mathbf{A} \right\| _2\left( \left\| \mathbf{x}_i \right\| _2+\left\| \mathbf{x}_{i-1} \right\| _2 \right). 
\end{align}
Then using \eqref{eq: temp_x}, we have
\begin{align}
     \alpha _{i-1}{\bf A}\mathbf{p}_{i-1}={\bf A} \left(\mathbf{x}_i-\mathbf{x}_{i-1}-\boldsymbol{\xi }_i\right), \label{eq: temp_x2},
\end{align}
and substituting \eqref{eq: temp_x2} into the second term in \eqref{eq: eta}, we have
\begin{align}
    \tau _{\mathrm{W}}\left\| \alpha _{i-1}\mathbf{Ap}_{i-1} \right\| _2\le \left( u_{\rm W}+\sqrt{2}\gamma _{2}^{\left( \rm W \right)}+\mathcal{O} \left( u_{\mathrm{W}}^{2} \right) \right) \varXi _i. \label{eq: eta2}
\end{align}
Furthermore, assume that each term $\left\| \boldsymbol{\eta }_k \right\| _2$, $k=1,\cdots ,i-1$ is bounded by $\mathcal{O} \left( u_{\rm W}+u_{\mathrm{MV}} \right) \left\| \mathbf{A} \right\| _2\left( \left\| \mathbf{x} \right\| _2+\max_{\ell \le k} \left\| \mathbf{x}_{\ell} \right\| _2 \right) $. It is evident that $\boldsymbol{\eta }_1$ satisfies the bound. Since ${\bf r}_{i-1}$ satisfies
\begin{align}
    \mathbf{r}_{i-1}=\mathbf{b}-\mathbf{Ax}_{i-1}-\left( \mathbf{b}-\mathbf{Ax}_0-\mathbf{r}_0 \right) +\sum_{j=1}^{i-1}{\left( \mathbf{A}\boldsymbol{\xi }_j+\boldsymbol{\eta }_j \right)} \notag.
\end{align}
Using \eqref{eq: first_term} and \eqref{eq: second_term}, we have
\begin{align}
    &\left\| \mathbf{r}_{i-1} \right\| _2\le \left\| \mathbf{A} \right\| _2\left\| \mathbf{x}-\mathbf{x}_{i-1} \right\| _2\notag\\
    &+\left\| \mathbf{A} \right\| _2\left\| \mathbf{x} \right\| _2\left( 3u_{\rm W}+2\sqrt{2}\gamma _{2}^{\left( \rm W \right)}+\mathcal{O} \left( u_{\rm W}^{2} \right) \right) \left( i-1 \right) \Theta _i \notag \\
    & +\left( i-1 \right) \mathcal{O} \left( u_{\rm W}+u_{\mathrm{MV}} \right) \left\| \mathbf{A} \right\| _2\left( \left\| \mathbf{x} \right\| _2+\max_{\ell \le i-1} \left\| \mathbf{x}_{\ell} \right\| _2 \right) \label{eq: eta1}
\end{align}
Substituting \eqref{eq: eta3}, \eqref{eq: eta2}, and \eqref{eq: eta1} into the bound \eqref{eq: eta}, we obtain
\begin{align}
    \left\| \boldsymbol{\eta }_i \right\| _2&\le \left( u_{\mathrm{W}}+\mathcal{O} \left( u_{\mathrm{W}}u_{\mathrm{MV}} \right) \right) \left\| \mathbf{A} \right\| _2\left\| \mathbf{x} \right\| _2 \notag \\
    &+\mu _{\mathrm{W},\mathrm{MV}}\left\| \mathbf{A} \right\| _2\max_{\ell \le i} \left\| \mathbf{x}_{\ell} \right\| _2,\label{eq: temp_eta}\\
    \mu _{\mathrm{W},\mathrm{MV}}&=3u_{\mathrm{W}}+2\sqrt{2}\gamma _{2}^{\left( \mathrm{W} \right)}+2\sqrt{2N}\gamma _{2N}^{\left( \mathrm{MV} \right)}+\mathcal{O} \left( u_{\mathrm{W}}u_{\mathrm{MV}} \right) \notag.
\end{align}
\eqref{eq: temp_eta} shows that $\left\| \boldsymbol{\eta }_i \right\| _2$ is also bounded by $\mathcal{O} \left( u_{\rm W}+u_{\mathrm{MV}} \right) \left\| \mathbf{A} \right\| _2\left( \left\| \mathbf{x} \right\| _2+\max_{\ell \le i} \left\| \mathbf{x}_{\ell} \right\| _2 \right) $, so the induction is complete and \eqref{eq: temp_eta} is proved. Based on \eqref{eq: temp_eta}, the right third term in \eqref{eq: b-ax} can be expressed as
\begin{align}
    \sum_{j=1}^i{\frac{\left\| \boldsymbol{\eta }_j \right\| _2}{\left\| \mathbf{A} \right\| _2\left\| \mathbf{x} \right\| _2}}\le i\left[ u_{\mathrm{W}}+\mathcal{O} \left( u_{\mathrm{W}}u_{\mathrm{MV}} \right) +\mu _{\mathrm{W},\mathrm{MV}}\Theta _i \right]. \label{eq: third_term}
\end{align}

Finally, substituting the bounds \eqref{eq: first_term}, \eqref{eq: second_term}, and \eqref{eq: third_term} into \eqref{eq: all}, and neglecting the high order term of $u_{\rm MV}$ based on \eqref{eq: gamma}, we have
\begin{align}\label{eq: final}
    \frac{\left\| \mathbf{b}-\mathbf{Ax}_i-\mathbf{r}_i \right\| _2}{\left\| \mathbf{A} \right\| _2\left\| \mathbf{x} \right\| _2}\le \left( 1+i \right) u_{\mathrm{W}}+ \varUpsilon_i\Theta _i, 
\end{align}
where 
\begin{align}
    \varUpsilon_i = \left( \left( 8\sqrt{2}+6 \right) i+1 \right) u_{\mathrm{W}}+\left( 2i+1 \right) 2\sqrt{2}N^{\frac{3}{2}}u_{\mathrm{MV}}\notag.
\end{align}

When \textit{Algorithm} \ref{alg: FP-CG} converges, we have \cite{greenbaum1997estimating}
\begin{align}
    \left\| \mathbf{A}^{1/2}\left( \mathbf{x}-\mathbf{x}_k \right) \right\| _2\le \left\| \mathbf{A}^{1/2}\left( \mathbf{x}-\mathbf{x}_0 \right) \right\| _2, \label{eq: con}
\end{align}
Using the Rayleigh principle and \eqref{eq: con} gives
\begin{align}
    \sqrt{\lambda _{\min}\left( \mathbf{A} \right)}\left\| \mathbf{x}-\mathbf{x}_k \right\| _2\le \left\| \mathbf{A}^{1/2}\left( \mathbf{x}-\mathbf{x}_k \right) \right\| _2\\
    \le \left\| \mathbf{A}^{1/2}\left( \mathbf{x}-\mathbf{x}_0 \right) \right\| _2\le \sqrt{\lambda _{\max}\left( \mathbf{A} \right)}\left\| \mathbf{x}-\mathbf{x}_0 \right\| _2,
\end{align}
Hence, we have
\begin{align}
    &\left\| \mathbf{x}-\mathbf{x}_k \right\| _2\le \sqrt{\kappa \left( \mathbf{A} \right)}\left\| \mathbf{x}-\mathbf{x}_0 \right\| _2 \\
\Longrightarrow &\left\| \mathbf{x}_k \right\| _2=\left\| \mathbf{x}-\left( \mathbf{x}-\mathbf{x}_k \right) \right\| _2\notag \\
&\le \left\| \mathbf{x} \right\| _2+\left\| \mathbf{x}-\mathbf{x}_k \right\| _2 \notag \\
&\le \left\| \mathbf{x} \right\| _2+\sqrt{\kappa \left( \mathbf{A} \right)}\left( \left\| \mathbf{x} \right\| _2+\left\| \mathbf{x}_0 \right\| _2 \right) 
\end{align}
Further, based on the definition of $ \Theta _i$ and the initial value of ${\bf x}_0 = {\bf 0}$, we give
\begin{align}
    \Theta _i&=\max_{k\le i} \frac{\left\| \mathbf{x}_k \right\| _2}{\left\| \mathbf{x} \right\| _2}\le 1+\sqrt{\kappa \left( \mathbf{A} \right)}\left( 1+\Theta _0 \right) \notag \\
    &=1+\sqrt{\kappa \left( \mathbf{A} \right)}, \label{eq: theta_condition}
\end{align}
Substituting \eqref{eq: theta_condition} into \eqref{eq: final} gives 
\begin{align}
    \frac{\left\| \mathbf{b}-\mathbf{Ax}_i-\mathbf{r}_i \right\| _2}{\left\| \mathbf{A} \right\| _2\left\| \mathbf{x} \right\| _2}&\le \left( 1+i \right) u_{\mathrm{W}}+\varUpsilon_i \left( 1+\sqrt{\kappa \left( \mathbf{A} \right)} \right) \\
     &\sim \mathcal{O} \left( N^{\frac{3}{2}}u_{\mathrm{MV}}\sqrt{\kappa \left( \mathbf{A} \right)} \right).
\end{align}

Hence, \textit{Theorem} \ref{the: Attainable} holds.

\section{Proof of \textit{Theorem} \ref{the: correlated}}
\label{app: correlated}
For simplification, we denote ${\bf V} = \sqrt{\mathbf{R}_r}$ and ${\bf U} = \sqrt{\mathbf{R}_t}$ with ${\bf V}^\mathsf{T}={\bf V}$ and ${\bf U}^\mathsf{T}={\bf U}$, respectively. Then, the $(m,n)$-th element of $\bf H$ is given by
\begin{align}
    h_{mn}=\sum_{a=1}^M{\sum_{b=1}^N{v_{ma}\omega _{ab}u_{bn}}},
\end{align}
where $v_{ma}$ is the $(m,a)$-th element of $\bf V$ and $u_{bn}$ is the $(b,n)$-th element of $\bf U$. Since $h_{mn}$ is linear combination of a series of i.i.d complex Gaussian variables $\omega _{ab}$ with $a=1,\cdots,M$ and $b=1,\cdots,N$, $h_{mn}$ is also a complex Gaussian variable. And the expectation and variance of $h_{mn}$ can be expressed as
\begin{align}
    \mathbb{E} \left\{ h_{mn} \right\} &=0, \\
    \mathbb{V} \left\{ h_{mn} \right\} &=\mathbb{E} \left\{ \left| h_{mn} \right|^2 \right\} =\mathbb{E} \left\{ h_{mn}h_{mn}^{*} \right\} \notag \\
    &=\mathbb{E} \left\{ \sum_{a,b,k,l}{v_{ma}\omega _{ab}u_{bn}v_{mk}\omega _{kl}^{*}u_{ln}} \right\} \notag \\
    &=\frac{1}{M}\sum_{a=1}^M{\sum_{b=1}^N{v_{ma}^{2}u_{bn}^{2}}} \notag \\
    & =\frac{1}{M}\left[ \mathbf{R}_r \right] _{mm}\left[ \mathbf{R}_t \right] _{nn}=\frac{1}{M}.\label{eq: var_hmn}
\end{align}

More general, the expectation of $h_{mn}^{*}h_{ji}$ is given by
\begin{align}
    \mathbb{E} \left\{ h_{mn}^{*}h_{ji} \right\} &=\mathbb{E} \left( \sum_{k=1}^M{\sum_{l=1}^N{v_{mk}\omega _{kl}^{*}u_{ln}}} \right) \left( \sum_{c=1}^M{\sum_{d=1}^N{v_{jc}\omega _{cd}u_{di}}} \right) \notag \\
    &=\frac{1}{M}\sum_{k=1}^M{\sum_{l=1}^N{v_{mk}v_{jk}u_{ln}u_{li}}} \notag \\
    &=\frac{1}{M}\left[ \mathbf{R}_r \right] _{mj}\left[ \mathbf{R}_t \right] _{ni}. \label{eq: h_mnh_mi}
\end{align}

Further, the covariance of $h_{mn}^{*}h_{mi}$ and $h_{on}^{*}h_{oi}$ can be expressed as
\begin{align}
    &~~~\,\mathbb{C} \left( h_{mn}^{*}h_{mi},h_{on}^{*}h_{oi} \right) \notag \\
    &=\mathbb{E} \left( h_{mn}^{*}h_{mi}h_{on}h_{oi}^{*} \right) 
    -\mathbb{E} \left( h_{mn}^{*}h_{mi} \right) \mathbb{E} \left( h_{on}h_{oi}^{*} \right) \notag \\
    &\overset{\left( a \right)}{=}\mathbb{E} \left( h_{mn}^{*}h_{mi} \right) \mathbb{E} \left( h_{on}h_{oi}^{*} \right) +\mathbb{E} \left( h_{mn}^{*}h_{on} \right) \mathbb{E} \left( h_{mi}h_{oi}^{*} \right)\notag \\
    &-\mathbb{E} \left( h_{mn}^{*}h_{mi} \right) \mathbb{E} \left( h_{on}h_{oi}^{*} \right) \notag \\
    &=\mathbb{E} \left( h_{mn}^{*}h_{on} \right) \mathbb{E} \left( h_{mi}h_{oi}^{*} \right) \notag \\
    &=\frac{1}{M^2}\left[ \mathbf{R}_r \right] _{mo}^{2}, \label{eq: cov}
\end{align}
where $(a)$ uses the Isserlis’ theorem for complex Gaussian variables \cite{isserlis1918formula,1057719}.

Based on \eqref{eq: var_hmn}, \eqref{eq: h_mnh_mi} and \eqref{eq: cov}, we have
\begin{align}
    &\mathbb{V} \left\{ \mathbf{h}_{n}^{\mathsf{H}}\mathbf{h}_i-\sum_{m=1}^M{\mathbb{E} \left\{ h_{mn}^{*}h_{mi} \right\}} \right\} =\mathbb{V} \left\{ \sum_{m=1}^M{h_{mn}^{*}h_{mi}} \right\} \notag \\
    &=\sum_{m=1}^M{\mathbb{V} \left\{ h_{mn}^{*}h_{mi} \right\}}+2\sum_{m=1}^{M-1}{\sum_{o=m+1}^M{\mathbb{C} \left( h_{mn}^{*}h_{mi},h_{on}^{*}h_{oi} \right)}}\notag\\
    &=\sum_{m=1}^M{\left( \mathbb{E} \left\{ \left| h_{mn}^{*}h_{mi} \right|^2 \right\}-\left| \mathbb{E} \left\{ h_{mn}^{*}h_{mi} \right\} \right|^2 \right)} \notag \\
    &+2\sum_{m=1}^{M-1}{\sum_{o=m+1}^M{\mathbb{C} \left( h_{mn}^{*}h_{mi},h_{on}^{*}h_{oi} \right)}}\notag\\
    &\overset{\left( a \right)}{=}\sum_{m=1}^M{\mathbb{E} \left\{ \left| h_{mn} \right|^2 \right\} \mathbb{E} \left\{ \left| h_{mi} \right|^2 \right\}}+\frac{2}{M^2}\sum_{m=1}^{M-1}{\sum_{o=m+1}^M{\left[ \mathbf{R}_r \right] _{mo}^{2}}}\notag\\
    &=\frac{1}{M}+\frac{2}{M^2}\left[ \varsigma \left( M-1 \right) -\varsigma ^2\left( 1-\zeta ^{2\left( M-1 \right)} \right) \right], \label{eq: var}
\end{align}
where $\mathbf{h}_{n}$ and $\mathbf{h}_{i}$ is the $n$-th and $i$-th column of ${\bf H}$ with $1\leq n,i\leq N$, and
\begin{align}
    \varsigma =\frac{\zeta ^2}{1-\zeta ^2}.
\end{align}

Finally, from \eqref{eq: var} and $0\leq \zeta < 1$, when $M$ is asymptotically large, i.e., $M\longrightarrow \infty$, we obtain
\begin{align}
    \mathbb{V} \left\{ \mathbf{h}_{n}^{\mathsf{H}}\mathbf{h}_i-\sum_{m=1}^M{\mathbb{E} \left\{ h_{mn}^{*}h_{mi} \right\}} \right\}\longrightarrow 0.
\end{align}
Hence, we can further have
\begin{align}
    &\mathbf{h}_{n}^{\mathsf{H}}\mathbf{h}_i-\sum_{m=1}^M{\mathbb{E} \left\{ h_{mn}^{*}h_{mi} \right\}}\xrightarrow{a. s.}0,\\
    \overset{\eqref{eq: h_mnh_mi}}{\Longrightarrow}  ~&\mathbf{h}_{n}^{\mathsf{H}}\mathbf{h}_i\xrightarrow{a. s.}\left[ \mathbf{R}_t \right] _{ni},\\
    \Longrightarrow ~&\mathbf{H}^{\mathsf{H}}\mathbf{H}\xrightarrow{a. s.}\mathbf{R}_t.
\end{align}
Overall, the proof ends.

\section{Comparison between BJ preconditioning and other preconditioning methods}
\label{app: pre}
In this appendix, we compare the advantages of the BJ preconditioner with both traditional preconditioning methods, such as the IC approach \cite{7868573}, and more recent methods, such as the UW-SVD preconditioner \cite{10653741}. The comparison is conducted from two perspectives: convergence performance and computational complexity.

\subsubsection{Convergence Performance} 
\begin{figure}[t]
     \centering
     \includegraphics[width=0.4\textwidth]{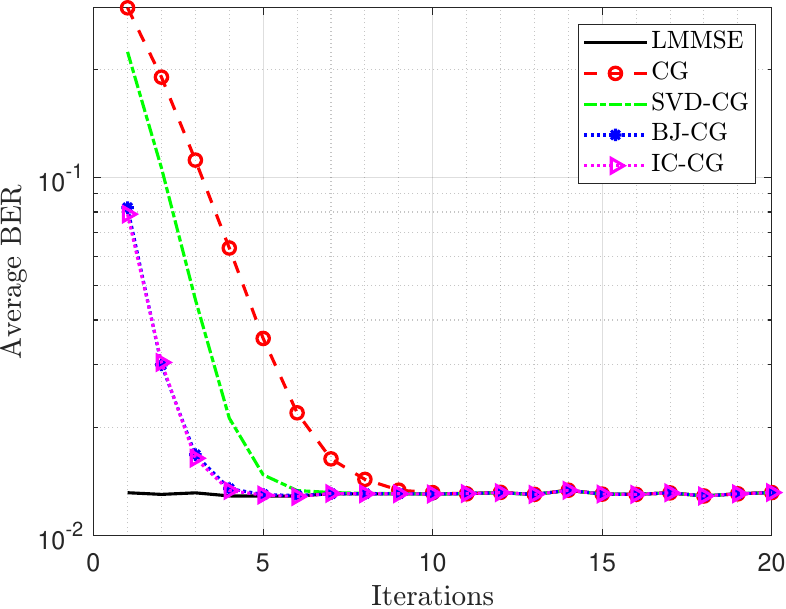}
     \caption{Convergence curve of different preconditioning CG detection with $\zeta = 0.5$ in SNR $=20$ dB.}
     \label{fig: precondition}
    \end{figure}

Fig. \ref{fig: precondition} compares the convergence behavior of BJ-, IC-, and UW-SVD-based CG detection under $\zeta = 0.5$ at ${\rm SNR}=20$ dB. It can be observed that both BJ-CG and IC-CG achieve faster convergence than UW-SVD-CG detection.  

\subsubsection{Computational Complexity} Table \ref{tab: pre} summarizes the computational complexity of generating different preconditioning matrices. The IC method exhibits an asymptotic complexity of $\mathcal{O}(N^3)$, while the UW-SVD method scales linearly with the number of BS antennas $M$. In contrast, the BJ method requires substantially lower complexity, on the order of $\mathcal{O}(NL^2)$, and is independent of $M$, thereby offering a favorable trade-off between performance and efficiency.  
    
    \begin{table*}[t]
    \centering 
    \setlength{\tabcolsep}{3.5pt}
    \caption{Computational Complexity of generating the different preconditioning matrices}
    \label{tab: pre}
    \begin{threeparttable}
    \begin{tabular}{cccc}
    \toprule[1pt]
    \midrule
    Preconditioning methods& BJ & IC\tnote{(1)} & UW-SVD  \\ \midrule
    Computational Complexity   &  $\mathcal{O}\left(NL^2\right)$    & $\mathcal{O}\left(\left(N^2-2S\right)^{3/2}-\left(N^2-2S\right)^{1/2}\right)$ & $\mathcal{O}\left(N_{\mathrm{UE}}MN\right)$\\ \midrule
    \bottomrule[1pt]
    \end{tabular}       
    \begin{tablenotes}    
        \footnotesize               
        \item[(1)] $S$ is the number of zero entries.          
    \end{tablenotes} 
    \end{threeparttable}
    \end{table*} 
Overall, the BJ preconditioner achieves competitive convergence performance with significantly lower computational complexity compared to both IC and UW-SVD methods, making it particularly well-suited for finite-precision massive MIMO detection.

\section{Proof of \textit{Theorem} \ref{the: compare_Condition}}
\label{app: compare_Condition}
We use ${\bf A}_{\rm LMMSE}$ as an example. The results can be easily reduced to that of ${\bf A}_{\rm ZF}$. Specifically, the matrix ${\bf A}_{\rm LMMSE}$ can be expressed as 
\begin{align}
    {\bf A}_{\rm LMMSE}=\left[ \mathbf{H}_1,\cdots, \mathbf{H}_K \right]^\mathsf{H}\left[ \mathbf{H}_1,\cdots, \mathbf{H}_K \right] + \sigma_n^2 {\bf I}_N.\label{eq: A_mmse_1}
\end{align}
Using \textit{Assumption} \ref{ass: h}, \eqref{eq: A_mmse_1} can be further simplified as
\begin{align}
    \mathbf{A}_{\mathrm{LMMSE}}=\left[ \begin{matrix}
	\mathbf{H}_{1}^{\mathsf{H}}\mathbf{H}_1+\sigma _{n}^{2}\mathbf{I}_{N_{\mathrm{UE}}}&		&		\\
	&		\ddots&		\\
	&		&		\mathbf{H}_{K}^{\mathsf{H}}\mathbf{H}_K+\sigma _{n}^{2}\mathbf{I}_{N_{\mathrm{UE}}}\\
\end{matrix} \right]. \label{eq: A_mmse_2}
\end{align}

It is evident that \eqref{eq: A_mmse_2} is a block diagonal matrix satisfying
\begin{align}
    \kappa\left({\bf A}_{\rm LMMSE}\right)\geq \max_{1\leq k\leq K} \kappa\left(\mathbf{H}_k^\mathsf{H}\mathbf{H}_k + \sigma_n^2 {\bf I}_{N_{\rm UE}}\right).
\end{align}
Note that the intra-user channel columns are correlated, and then $\kappa\left(\mathbf{H}_k^\mathsf{H}\mathbf{H}_k + \sigma_n^2 {\bf I}_{N_{\rm UE}}\right)>1$, i.e.,
\begin{align}
    \kappa\left({\bf A}_{\rm LMMSE}\right) >1. \label{eq: use_1}
\end{align}

Furthermore, since ${\bf A}_{\rm LMMSE}$ in \eqref{eq: A_mmse_2} is a block diagonal matrix, for BJ precondition matrix in \eqref{eq: pre_BJ} with $L\geq N_{\rm UE}$, we have
\begin{align}
    \Bar{\bf A} = {\bf M}_d^{1/2}{\bf A}{\bf M}_d^{1/2} = {\bf I}_N,
\end{align}
which indicates that the condition number of $\Bar{\bf A}$ is 1. Finally, using \eqref{eq: use_1}, \eqref{eq: convergence} in \textit{Theorem} \ref{the: compare_Condition} holds.

{
\section{Computation Complexity Comparison at high SNR Levels} \label{app: high_snr}
To further show the advantage of the proposed methods in terms of computational complexity, we have performed a target-BER comparison. Specifically, we fix a target BER equal to the BER achieved by LMMSE at SNR = 25 dB. For each detection method, we then measure the computational complexity (in terms of number of multiplications) required to reach that target BER at the same SNR (25 dB). These results are shown in Figs.~\ref{fig: ber25} and \ref{fig: cc25}. It can be observed that achieving LMMSE-equivalent BER requires 17, 21, and 13 iterations for CG, FP-CG, and FP-BJ-CG detection, respectively. Moreover, the proposed FP-BJ-CG method with $\mathtt{fp16}$ reduces computational complexity by $50.0\%$ compared to CG with $\mathtt{fp64}$, and by $73.4\%$compared to direct LMMSE inversion with $\mathtt{fp64}$, while achieving identical BER. For FP-CG with $\mathtt{fp32}$, the reductions are $38.2\%$ and $67.2\%$, respectively.

 \begin{figure}[t]
     \centering
     \includegraphics[width=0.4\textwidth]{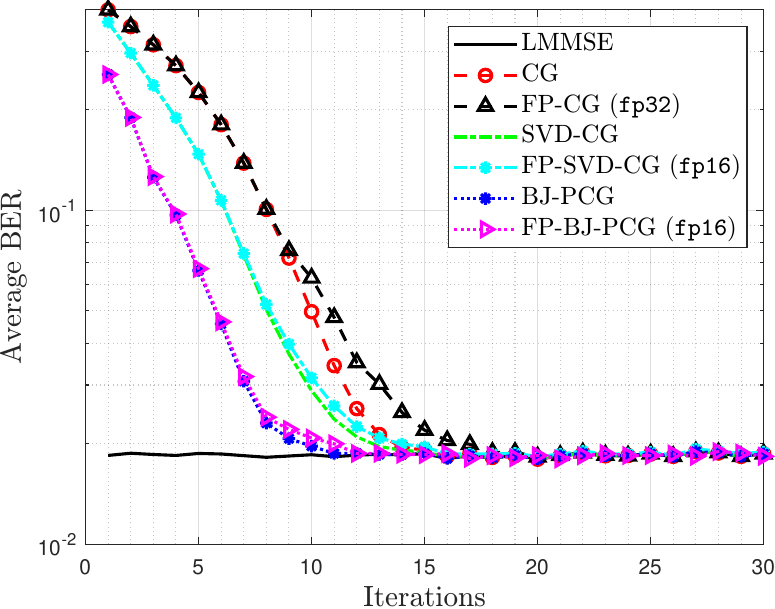}
     \caption{{Convergence curve of various detection using the heuristic method with $\zeta = 0.8$ in SNR $=25$ dB.}}
     \label{fig: ber25}
 \end{figure}

  \begin{figure}[t]
     \centering
     \includegraphics[width=0.4\textwidth]{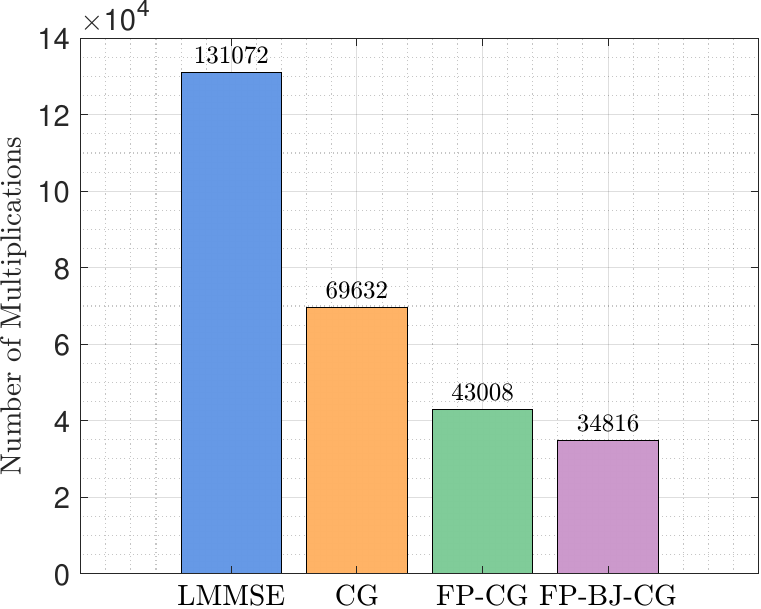}
     \caption{{Computational complexity of different detection methods required to achieve the target BER equivalent to LMMSE performance with $\zeta = 0.8$ in SNR $=25$ dB.}}
     \label{fig: cc25}
 \end{figure}
}

\bibliographystyle{IEEEtran}
\bibliography{reference}

\end{document}